\documentstyle[12pt]{article}
\begin{document}
\title{Nonaxial octupole deformations in
       light $N=Z$ nuclei at high spins}
\author{Takeshi Tanaka and Rashid G. Nazmitdinov\\
Bogoliubov Laboratory of Theoretical Physics, 
Joint Institute for Nuclear Research\\
141980 Dubna Moscow Region, Russia.\\
Kazuo Iwasawa\\
Information Processing Center, Hiroshima University\\
Hiroshima 739-8526, Japan.}
\date{9 Apr. 2000}

\maketitle
\newcommand{\hh} { \hat{H}    }
\newcommand{\hr} { \hat{R}    }
\newcommand{\hjz}{ \hat{J}_z  }
\newcommand{\hjy}{ \hat{J}_y  }
\newcommand{\hz} { \hat{Z}    }
\newcommand{\hn} { \hat{N}    }
\newcommand{\hq} { \hat{Q}}
\newcommand{\ha} { \hat{\alpha}         }
\newcommand{\had}{ \hat{\alpha}^\dagger }
\newcommand{\hc} { \hat{c}         }
\newcommand{\hcd}{ \hat{c}^\dagger }
\newcommand{\hP} { \hat{P}    }
\newcommand{\mr} {\mbox{\boldmath$\rho$}}
\newcommand{\mj} {\mbox{\boldmath$j$}}
\newcommand{\mq} {\mbox{\boldmath$q$}}
\newcommand{\bkO}  {\mbox{\footnotesize $(\omega)$}}
\newcommand{\bkOz} {\mbox{\footnotesize $(\omega=0)$}}
\newcommand{\bkOc} {\mbox{\footnotesize $(\omega_c)$}}
\newcommand{\moist}{ {\cal J}^{\rm (1)} }
\newcommand{\moi}  { {\cal J}^{\rm (2)} }
\newcommand{\moiing} { {\cal J}_{\rm IB} }
\newcommand{\lang}{\langle}
\newcommand{\rang}{\rangle}
\newcommand{\brao}{\lang \phi\bkO|}
\newcommand{\keto}{|\phi\bkO\rang}
\newcommand{\ketoz}{|\phi\bkOz\rang}
\newcommand{\ketoc}{|\phi\bkOc\rang}
\begin{abstract}
High spin states of $^{32}S$ and $^{56}Ni$ are investigated
by means of the cranking Hartree-Fock method with the Gogny
interaction without imposing a restriction on
the axial reflection symmetry.
It was found that a non-axial octupole deformation of
the $Y_{31}$ type becomes important in the yrast states
of $^{32}S$. A similar effect is predicted for the nucleus
$^{56}Ni$.
\end{abstract}

PACS number(s): 21.60-n; 21.60.Jz; 27.30.+t; 27.40.+z
\section{Introduction}
\label{sec:intro}

One of the most important concepts in the
many--body theory of finite Fermi systems
is the mean field approach.
In fact, many phenomena  observed in
nuclei can be explained by means of a
spontaneous symmetry breaking mechanism which leads
to a mean field solution that does not obey symmetries of
the original many--body Hamiltonian \cite{BM75,RS}.
A nucleus with a partially filled highest shell spontaneously
deforms in such a way that a lower energy minimum is achieved,
and thus a non--spherical equilibrium shape is attained.
Coriolis forces may cause a similar  effect for a
nucleus which has a spherical shape in a ground state.
In particular, a strong coupling of normal and intruder
states near the Fermi surface at large rotational frequencies
can lead to a superdeformed or
to an octupole deformed shape for certain combinations of
protons and neutrons \cite{Ab}.
The most practical method for the analysis of nuclear shapes 
is a phenomenological macroscopic--microscopic method (MMM)
which combines the liquid drop model describing the bulk properties of
nuclear matter and the Strutinsky Shell Correction method
providing the description of quantum shell effects of
phenomenological potentials \cite{RS}.
The more fundamental approach is
based on self--consistent Hartree-Fock (HF) calculations
once a particular choice of nucleon-nucleon interaction has been
made. A commonly used inter-nucleon force is that of Skyrme.
While the HF+Skyrme approach describes  major features of nuclei
quite well,
it does not completely take into account the pairing correlations.
In addition, various sets of parameters for Skyrme forces
may not provide a definite answer in some cases.
The Hartree-Fock-Bogoluibov (HFB) method with the Gogny forces
resolves these problems quite effectively.
Furthermore, it has the same prediction power as the MMM \cite{col}.

In recent years a lot of experimental and theoretical efforts
has been devoted to the analysis of superdeformed (SD) bands in
different mass region \cite{Ab,Nol,Jan,Bak,Bak1}.
On the other hand,
the study of octupole degrees of freedom is also
a topical subject in the nuclear structure physics \cite{But,Frau}.
It turns out that octupole deformations are significant
for superdeformed nuclei as well as for normal-deformed (ND) nuclei.
Most of these studies were restricted to the axial octupole
deformation proportional to the $Y_{30}$ term in the octupole
family. The reason for this is  primarily  computational
complications arising as a result of the extra degree of freedom
introduced by non-axial octupole deformations.
However, during the last few years some remarkable results
have been reported.
It was found that $Y_{31}$ and $Y_{32}$
components resulted in the lowest energy octupole vibrations
for oblate and prolate superdeformed shapes and therefore may
have consequences for octupole instability \cite{Nak,RN,But}.
Calculations using the MMM with the Woods--Saxon potential
predicted the importance of the banana-type $Y_{31}$
deformation for highly deformed nuclei \cite{Ch},
while the $Y_{32}$ deformation has been found to be important
in the $^{222}Ra$ nucleus \cite{Dud}.
The self-consistent HF+Skyrme calculations  \cite{Tak98}
suggest the softness of the oblate states in $A\sim80$ nuclei
against the $Y_{33}$ deformation in the ground states. The
self-consistent cranking HF+Skyrme approach \cite{Yam98} predicts
that the $Y_{31}$ deformation is important for a correct description
of the yrast  band of $^{32}S$. Moreover,
the study of octupole deformations sheds light
on the interplay between a classical chaos and a quantum spectrum
of finite Fermi systems \cite{prl,mat,He98}.
Guided by thorough investigation of non-axial octupole deformations
in a harmonic oscillator model \cite{He99}, we aim to
analyze how symmetries break at high spins in
the cranking  HFB approach with the Gogny interaction.
In accordance with  the octupole instability suggested
at particle number $N=$16 and 28 \cite{He99},
we choose two $N=Z$ nuclei, $^{32}S$ and $^{56}Ni$.
In Sec.\ref{sec:HFB} we review the main features of our
model. The discussion of main results is presented
in Sec.\ref{sec:num} with the conclusion following in
Sec.4. In the Appendix we introduce a simple two-level model
to understand qualitatively  the behavior of
the Inglis-Beliaev moment of inertia due to strong octupole
coupling between two single-particle (s.p.) states.

\section{The Model}
\label{sec:HFB}

As mentioned above, the present  self-consistent cranking
HFB calculation has been performed with
the effective Gogny D1S interaction
\cite{Gog75,Dev80,Gir83,Ber84}. This interaction
provides a good description of many nuclear properties over
the nuclide chart, e.g.,
ground state energies, odd-even energy differences,
electron scattering data \cite{Dev80},
fission barriers \cite{Ber84}.
Also, a good description is obtained for
bulk properties
of rotating nuclei in actinide region \cite{Egi93},
mercury region \cite{Gir94} and fp-shell region \cite{Cau95}.
Our numerical code  has been
used for the analysis of microscopic dynamics in light rotating
nuclei \cite{Tan98,Tan97,Iwa97}.
 In these calculations, however,
the signature symmetry was conserved (see discussion
about different symmetries in rotating nuclei in
\cite{Frau,Dob99}). The present calculations are performed without
assuming {\it {\'a} priori} the axial and signature symmetries.
In addition, our Hamiltonian includes the Coulomb interaction
and the center of mass correction up to the exchange terms exactly.

The $z$-axis is taken as a rotational axis in our code.
To save the CPU time in numerical calculations, we impose the
$\hP e^{-i\pi \hjz}$ ($z$-simplex)
and the $\hP e^{-i\pi \hjy} \hat{\tau}$ ($\hat{S}^T_y$)
symmetries \cite{Dob99},
where $\hP$ is the parity operator,
$e^{-i\pi \hat{J}_i}$ is the rotation operator
around the $i$-axis (i=y,z) on angle  $\pi$ ,
and $\hat{\tau}$ is the time-reversal operator.
Due to the $z$-simplex and $\hat{S}^T_y$ symmetries,
the mass asymmetry of nucleus is allowed only along the $x$-axis.
Thus, we solve numerically the following
cranking HFB equations:
{\setcounter{enumi}{\value{equation}}
\addtocounter{enumi}{1}
\setcounter{equation}{0}
\renewcommand{\theequation}{\theenumi\alph{equation}}
\begin{eqnarray}
  \label{crhfbeqa1}
  \delta \brao \hh - \lambda_p \hz
                   - \lambda_n \hn
                   - \omega    \hjz&&\nonumber \\
                   + \mu_x \brao \hat{x} \keto \hat{x}\keto &=& 0.\\
  \label{crhfbeqb1}
         \brao                 \hz  \keto  =  Z, \quad
         \brao                 \hn  \keto &=& N, \\
  \label{crhfbeqc1}
         \brao                 \hjz \keto &=& I,\\
  \label{comc}
         \brao             \hat{x} \keto &=& 0,
\end{eqnarray}
\setcounter{equation}{\value{enumi}}}
where the Lagrange multipliers $\lambda_p$ and $\lambda_n$
are the chemical potentials of proton and neutron fields, respectively
(the operators $\hz$ and $\hn$ are protons
and neutrons number operators);
the Lagrange multiplier $\omega$ is the angular frequency of a
collective rotation around the $z$-axis and
$\hjz$ is the $z$-component of the angular momentum operator
$\hat{\mbox{\boldmath$J$}}$.
To keep the center of mass motion fixed,
we also impose the quadrupole constraint operator
$\mu_x \brao \hat{x} \keto \hat{x}$ \cite{Flo73}
to the Routhian
$\hr = \hh - \lambda_p \hz - \lambda_n \hn - \omega    \hjz$
in the $x$-axis direction.

The quasi-particle (q.p.) operators
\begin{equation}
\had_i = \sum_k U_{ki} \hcd_k
                 + V_{ki}  \hc_k \quad ,
\end{equation}
where the state $|k\rang$ is a single-particle
basis state (see below),
obey the equation of motion
\begin{equation}
[\hr ,\had_i]=
\epsilon_i(\omega) \had_i
\end{equation}
which defines the quasi-particle energies $\epsilon_i(\omega)$
and quasi-particle amplitudes $U_{ki}$ and $V_{ki}$ as functions
of the rotational frequency $\omega$.

We define the $y$-axis as the quantization axis of deformation
for convenience.
Consequently, the quadrupole, $\beta_2$ and $\gamma$,
and octupole , $\beta_{3m}$, deformation parameters are
defined as
\begin{eqnarray}
\label{q2m}
\beta_{2}\cos \gamma &\equiv& \frac{4\pi}{5}
                    \frac{\lang r^2  Y_{2 0}(\theta, \varphi)  \rang}
                   {        AR_0^2}\\
       &=     & \sqrt{\frac{\pi}{ 5}}
            \frac{\lang  (3y^2 -r^2)\rang}
                   {        AR_0^2}, \nonumber\\
\beta_{2}\sin \gamma &\equiv& \frac{4\pi}{5}
                    \frac{\lang r^2 (Y_{2 2}(\theta, \varphi)
                                    +Y_{2-2}(\theta, \varphi) )\rang}
                   {\sqrt{2}AR_0^2}\\
       &=     & {\sqrt{\frac{3\pi}{ 5}}
            \frac{\lang  (x^2 - z^2)\rang}
                   {        AR_0^2},  }\nonumber
\end{eqnarray}
\begin{eqnarray}
\label{q3m}
\beta_{30} &\equiv& \frac{4\pi}{3}
                    \frac{\lang r^3  Y_{3 0}(\theta, \varphi)  \rang}
                   {        AR_0^3}\\
       &=     & \sqrt{\frac{7\pi}{ 9}}
            \frac{\lang y  (5y^2 -3r^2)\rang}
                   {        AR_0^3},\nonumber\\
\beta_{31} &\equiv& \frac{4\pi}{3}
                    \frac{\lang r^3 (Y_{3 1}(\theta, \varphi)
                                    -Y_{3-1}(\theta, \varphi) )\rang}
                   {\sqrt{2}AR_0^3}\\
       &=     &-\sqrt{\frac{21\pi}{18}}
            \frac{\lang x  (5y^2 - r^2 )\rang}
                   {        AR_0^3},\nonumber\\
\beta_{32} &\equiv& \frac{4\pi}{3}
                    \frac{\lang r^3 (Y_{3 2}(\theta, \varphi)
                                    +Y_{3-2}(\theta, \varphi) )\rang}
                   {\sqrt{2}AR_0^3}\\
       &=     & \sqrt{\frac{105\pi}{ 9}}
            \frac{\lang y  (x^2 - z^2)\rang}
                   {        AR_0^3},\nonumber\\
\beta_{33} &\equiv& \frac{4\pi}{3}
                    \frac{\lang r^3 (Y_{3 3}(\theta, \varphi)
                                    -Y_{3-3}(\theta, \varphi) )\rang}
                   {\sqrt{2}AR_0^3}\\
       &=     &-\sqrt{\frac{35\pi}{18}}
            \frac{\lang x   (x^2 -3z^2)\rang}
                   {        AR_0^3}, \nonumber
\end{eqnarray}
where $(r, \theta, \varphi)$ are spherical coordinates
related to the Cartesian coordinates in the rotating frame
$(x,y,z)$ as
\begin{equation}
\label{cood}
 (x,y,z) \equiv  ( r \sin \theta \cos \varphi,
                   r \cos \theta             ,
                   r \sin \theta \sin \varphi).
\end{equation}
All deformations proportional to the odd power of the $y$ coordinate
are forbidden in our code due to the  $\hat{S}^T_y$ symmetry.
Since $\beta_{30}$ and $\beta_{32}$ are
proportional to the odd $y^n$ terms,
we use $\beta_{31}$ and $\beta_{33}$
to represent the degree of the non-axial octupole deformation,
when the $y$-axis is the largest axis of a prolate nucleus.
Notice that the octupole deformation with a mass asymmetry
like a pear shape is also represented by the combination of
$\beta_{31}$ and $\beta_{33}$ when
the $x$-axis is the largest axis of a prolate
nucleus
\footnote{From these $\beta_{31}$ and $\beta_{33}$ values,
for an example,
we can obtain the value of {\bf $\beta_{30}$}
for the $x$-quantization axis as
\begin{equation}
\beta_{30}^x = \frac{\sqrt{6}}{4} \beta_{31}
             - \frac{\sqrt{10}}{4}\beta_{33}.
\end{equation}}.

The s.p. wave functions
have been expanded in a three-dimensional harmonic oscillator
basis up to the principal quantum number
$N=8$ for $^{32}S$ and up to $N=10$ for $^{56}Ni$.
The basis has been symmetrized with respect to the
$z$-simplex operation,
and eigenfunctions are eigenstates of the $\hat{S}^T_y$ operator
(the Goodman basis \cite{Goo79}).
Since the ground state shapes of the chosen nuclei
are a normal-deformed one for $^{32}S$ and a spherical one
for $^{56}Ni$,
we use a spherical Cartesian basis with the same
range parameters of the Hermite polynomials for all axes.
The range parameters have been optimized to reproduce
the largest binding energy of each ground state.

To understand the relation between s.p. degrees of freedom and
collective effects due to the rotation,
we calculate three moments of inertia,
the static moment of inertia $\moist\bkO\equiv I/\omega$,
the dynamical moment of inertia $\moi\bkO\equiv dI/d\omega$,
and the Inglis-Beliaev (IB) moment of inertia
\begin{equation}
  \label{thmoi}
  \moiing \bkO =
         2 \sum_{i>j}
  \frac{|J_{ij}\bkO|^2}
       {\epsilon_i       \bkO+\epsilon_j\bkO} ,
\end{equation}
which is the leading order term of the nuclear moment of inertia.
Here $J_{ij}\bkO$ is a matrix element of the angular momentum
operator $\hjz$
\begin{equation}
 \label{jij}
  J_{ij}\bkO = \brao [\ha_j\ha_i,\hjz]\keto.
\end{equation}
It is well known \cite{TV} that the IB moment of inertia
is too small to reproduce the absolute value of the
moment of inertia.
Comparing $\moiing \bkO$ and $\moist\bkO$ or $\moi\bkO$,
we can find the validity of using the $\moiing\bkO$ for
the analysis of  structural changes in the nuclear mean field.
In addition, we analyze different contributions
to the IB moment of inertia, Eq.(\ref{thmoi}),
\begin{equation}
\label{jingp1}
   \moiing\bkO = \sum_{\tau,s} {\cal J}_{\tau,s}\bkO \equiv
                   {\cal J}_{p,+1}\bkO + {\cal J}_{p,-1}\bkO
                 + {\cal J}_{n,+1}\bkO + {\cal J}_{n,-1}\bkO,
\end{equation}
where $(\tau,s)$ denote the isospin
$\tau=p,n$ and the $z$-simplex quantum number $s=\pm1$ which
characterize different subspaces.
The comparison of different components of $\moiing\bkO$
could provide the information
in which subspace the q.p. degrees of freedom
mainly affect the bulk properties of the nuclei.

The cranking HFB equations (\ref{crhfbeqa1}) are solved in
an iterative way.
As the convergence condition for each HFB state, we impose
the condition
\begin{equation}
\sum_i\left|\epsilon_i^{(n)}\bkO -
\epsilon_i^{(n-1)}\bkO\right|
\leq .1 \mbox{[KeV]},
\end{equation}
where $n$ is the number of iterations
in the course of solving the cranking HFB equations.

\section{Discussion of results}
\label{sec:num}
\subsection{$^{32}S$}

In Figs.1-3 results for the yrast line states,
angular momenta and deformations as functions of the rotational
frequency are presented.
The binding energy of the ground state in $^{32}S$
is well reproduced in our calculations
(Fig.\ref{fig:enes}).
We found that the contribution from the pairing interaction
terms to the total binding energy is almost negligible.
The energy gaps between q.p. energies near the Fermi  energies
are large, about 4 MeV, both in the proton and neutron fields.
Therefore, the calculations for the ND and SD bands
have been done using the cranking HF approximation with the same
parameter set as for the HFB calculations.
While there are differences in details between our calculations and
those  with  Skyrme forces \cite{Yam98},
we obtained  similar main results for the ND and SD bands.
The SD band becomes the yrast one for $I\geq 12 \hbar$.
The excitation energy of the SD minimum relative to the ground state
is about 10 MeV (Fig. \ref{fig:enes}).
Except for the binding energy,
these results are also consistent with those of SLy4-HF
calculations \cite{Mol99},
in which $\beta$-deformations are $\beta=.16$ and $\beta=.7$
in the ground state and SD minimum state,  respectively.
In comparison with the
results of the Skyrme III calculations \cite{Yam98},
however,
finite values of the non-axial octupole deformation $Y_{31}$
in the ND yrast band
are obtained at larger rotational frequency
$\omega\geq 1.5$ [MeV/ $\hbar$].
As is seen in Fig.\ref{fig:octs}(a),
the value of the non-axial octupole deformation $|\beta_{31}|$
increases suddenly at $\omega\sim 1.5$ [MeV/ $\hbar$],
where the pseudo level-crossing occurs between s.p. orbits
in the subspaces with $s=\pm 1$ (Fig.\ref{fig:sprs}(a)),
both in the proton and neutron fields.
These orbits can be associated
with the principal quantum numbers $N=2$ and $N=3$
at $\omega=0$[MeV/$\hbar$]. As a result, the
$Y_{31}$ deformation becomes favorable
at $\omega\sim 2.2$ [MeV/ $\hbar$], for $^{32}S$ with a finite value
$\beta_{31}=.13$. The density distribution projected on the
plane perpendicular to the rotation axis is shown on
Fig.\ref{fig:dens}(a) for the octupole band at
$\omega\sim 2.2$ [MeV$/ \hbar$] where the
$\beta_{31}$ deformation has a maximal value.
High resolution $\gamma -$ spectroscopy could
test the validity of different predictions based on different
nuclear interactions.

From the analysis of Figs.\ref{fig:octs}(a), \ref{fig:mois}(a)
it follows that
there is a correlation between the
behavior of the dynamical moment of inertia $\moi\bkO$
and the occurrence of the octupole deformation.
Similar to the behavior of the $|\beta_{31}|$,
the dynamical moment of inertia $\moi\bkO$ begins
to increase at $\omega\sim1.5$ [MeV/$\hbar$],
and both quantities show local maxima around
$\omega\sim 2$ [MeV/$\hbar$].
Due to the strong octupole interaction the quasi-crossing
between s.p. levels is smooth (Fig.\ref{fig:sprs}(a)).
The static moment of inertia $\moist\bkO$
is less sensitive to the quasi-crossing of s.p. levels,
and it changes smoothly with the increase of the
rotational frequency $\omega$.
However, the IB moment of inertia may reflect
the structural changes in a mean field like
the dynamical moment of inertia $\moi\bkO$.
In fact, it
also begins to increase at $\omega\sim1.5$ [MeV/$\hbar$]
and shows a local maximum at $\omega\sim 2.2$ [MeV/$\hbar$].
The correlation between the IB moment of inertia and the
magnitude of the octupole deformation $|\beta_{31}|$
may be understood within a simple two-level model
presented in the Appendix \ref{sec:octth}.
According to this model, the two rotation-aligned single-particle
states with a strong octupole
coupling near the Fermi-surface
may give a large contribution to the value of the IB moment
of inertia.
From Fig.\ref{fig:jinp}(a) follows that
the magnitudes of ${\cal J}_{p,+1}\bkO$ and ${\cal J}_{n,+1}\bkO$
(see Eq.(\ref{jingp1}))
are smaller than those from the $s=-1$ subspaces
which play the dominant roles in the collective rotation at low spins,
in the region $\omega<1.5$[MeV/$\hbar$]. However,
at the $\omega >1.5$[MeV/$\hbar$],
${\cal J}_{p,+1}\bkO$ and ${\cal J}_{n,+1}\bkO$
begin to increase rapidly,
and they show local maxima at $\omega\sim 2.2$[MeV/$\hbar$]
where $|\beta_{31}|$ also shows the maximum.
Since ${\cal J}_{p,-1}\bkO$ and ${\cal J}_{n,-1}\bkO$ are almost
unchanged in this region,
the occurrence of the $Y_{31}$ deformation
is mainly due to the s.p. quasi-crossing
in the $s=+1$ subspaces. To justify this statement
we calculate the quantity $\cos^2\psi\bkO$ (see Appendix).
Using the identity of the intruder orbits $\mu$ which have
odd parity at $\omega=0$[MeV/$\hbar$],
we evaluate the $\cos^2\psi\bkO$ approximately as:
\begin{equation}
\label{psi}
    \cos^2\psi\bkO \approx \sum_{n_x+n_y+n_z=odd}
    |\lang n_x,n_y,n_z ; \sigma |\mu\bkO\rang|^2,
\end{equation}
where a s.p. vector
$|n_x,n_y,n_z ; \sigma\rang$
is a component of the Goodman basis in the $s=+1$ subspace,
and $n_x$, $n_y$ and $n_z$ are the number of quanta
on the $x-$, $y-$ and $z-$ axes, respectively.
As is shown in Fig.\ref{fig:psi}(a),
the evaluated value of $\cos^2\psi\bkO = 1/2$ 
or equivalently $|\psi\bkO| = \pi/4$ 
at $\omega\sim 2.2$ [MeV$/ \hbar$]. 
According to the analysis within
the two-level model (see Appendix),
at this value of $|\psi\bkO|$
the mixing of s.p. states with opposite parities  is
the strongest one.

Similar to the results of \cite{Yam98}, our calculations
show rather shallow minima for the $\beta_{31}$ deformation.
In  Fig.\ref{fig:fluc}(a) we present results for
the fluctuations of octupole and
quadrupole deformations
\begin{equation}
\label{fluc1}
\Delta Q_{lm} \equiv
\frac{\sqrt{\brao \hq_{lm}^2 \keto -
            \brao \hq_{lm}   \keto^2}}
     {     |\brao \hq_{lm}   \keto  |}
\end{equation}
(see also Fig.\ref{fig:pesS}(a)). The contribution of the
ground state octupole correlations may improve the mean field results.
This contribution  could  be estimated,
for example, in the random phase approximation like \cite{KN}
according to the prescription \cite{DAN}, however,
this is beyond the scope of the present paper.

As is seen in Fig.\ref{fig:mois}(a),
both moments of inertia, $\moi\bkO$ and  $\moiing\bkO$,
begin to increase again at $\omega\sim 2.5$ [MeV$/ \hbar$] ,
which is apparently due to the quasi-crossings in the
$s=-1$ subspaces at this point.
There are also level-crossings between s.p. orbits
with different simplex numbers (Fig.\ref{fig:sprs}(a)),
which suggest the instability of the $z$-simplex symmetrized state
at high spins. We may speculate that the breaking of
the $z$-simplex symmetry due to the tilted or the
chiral rotation \cite{Frau} may lead to a better description
of high spin states in $^{32}S$ at $\omega\geq 2.5$ [MeV$/ \hbar$].

\subsection{$^{56}Ni$}

Our calculations reproduce
the binding energies and the charge radii of ground
states in  $^{56-60}Ni$   quite well (Fig. \ref{fig:nick}).
In this chain of nuclei we concentrate our attention on
$^{56}Ni$.

Though the shape of the ground state in $^{56}Ni$ is spherical,
we also found two shape isomers with the ND and SD
configuration at $\omega=0$ [MeV/$\hbar$],
which are $\sim 5$ MeV and $\sim 18$ MeV above
the ground state, respectively (see Fig.\ref{fig:eneni}).
The yrast non-rotating states observed in $^{56}Ni$ are related
to vibrational excitations. In fact, their properties can be studied
within the self-consistent cranking approach
with the rotation {\it around a symmetry axis} \cite{MN} and
we will discuss this subject in a forthcoming paper.
The calculations for the ground state,
ND  and SD bands  have been done
in a similar manner to those of
$^{32}S$ in the HF approximation, since
the contribution of the pairing interaction terms to the total binding
energies are almost negligible.
On Fig.\ref{fig:oms} the evolution of the angular momentum
with the increase of the rotational frequency is shown
for SD and ND bands.
At low spins the SD minimum is formed by
the 4$p$-4$h$ configuration with respect to
the ground state in $^{56}Ni$,
$\pi[(3)^{-2}(4)^{ 2}]\nu[(3)^{-2}(4)^{ 2}]$, which has a
prolate shape  with the quadrupole deformation $\beta=.6$
(Fig.\ref{fig:betas}).
This is also 4$h$ configuration with respect to the SD state
in $^{60}Zn$ due to the $N=30$ SD gap
in the $Zn$ isotopes \cite{Sve97},
which is also seen in Fig.\ref{fig:sprnisd}.
The MMM calculations \cite{She72,Ben84}
also predict the SD shape at large rotational frequencies.
We restricted, however, our analysis of the SD band in the region of
$\omega=0.0-0.4$[MeV/$\hbar$], since at larger rotational
frequencies the contribution of higher
shells $(N>10)$ becomes important.

Using the decomposition of the
rational harmonic oscillator (RHO) into the isotropic
ones, the relation between multi-clusters and mean field
results has been discussed  in \cite{ND}.
For SD shapes in the RHO  the one symmetric and one asymmetric combinations
of spherical oscillators are expected (see also \cite{RN}).
Within the RHO the SD states in $^{56}Ni$ correspond to
the asymmetric combination of two spherical oscillators
with magic numbers $40$ and $16$. 
The density distribution of the SD state
(see Fig.\ref{fig:sddni}) does not show such 
a multi-cluster structure at $\omega=0$ [MeV/$\hbar]$.
The possible reason is that our configuration space 
is too small. 
However, the SD minimum could be
related to the resonance
state in the $^{28}$Si+$^{28}$Si collision at high excitation energy
$E^\ast_{\rm CM}=65-70$MeV and high angular momenta
$I=34-42\hbar$ reported  in \cite{Bet81,Ueg94,Nou99}.
A thorough study of
$^{56}Ni$ as well as $^{32}S$ at high spins
may help to understand the link
between the tendency for nuclei to create
strongly deformed shapes and the tendency to develop
the cluster structure.

A prolate deformation $\beta=.35$ is found
for the ND minimum (Fig. \ref{fig:betas}) at $\omega= 0.0$ [MeV/$\hbar$].
With the increase of the rotational frequency the non-axial
octupole deformation $Y_{31}$ arises in the ND band.
As is seen in Fig.\ref{fig:octs}(b),
the value of the non-axial octupole deformation
increases rapidly around $\omega\sim .9$[MeV/$\hbar$].
It corresponds to $I\sim 10$[$\hbar$],
where the pseudo level-crossing occurs between s.p. orbits which
have the principal quantum numbers $N=3$ and $N=4$,
respectively, at $\omega=0$[MeV/$\hbar$] (Fig.\ref{fig:sprs}(b)).
The maximal value of the
$Y_{31}$ deformation, $\beta_{31}=.09$, is approached at
$\omega\sim 1.4$[MeV/$\hbar$] (Fig.\ref{fig:betas}(b)).
As is seen in Fig. \ref{fig:jinp}(b),
${\cal J}_{p,+1}\bkO$ and ${\cal J}_{n,+1}\bkO$
increase rapidly around $\omega\sim .9$[MeV/$\hbar$]
and approach the maximal value at $\omega\sim 1.3$[MeV/$\hbar$].
Using the two-level model and similar arguments as for the case
of $^{32}S$ we may conclude:
since ${\cal J}_{p,-1}\bkO$ and ${\cal J}_{n,-1}\bkO$
are decreasing in this region of the rotational
frequency, it is most likely that the octupole deformation is
caused by the quasi-crossing of s.p. levels,
which have different parities at $\omega=0$[MeV/$\hbar$],
in the $s=+1$ subspaces.
Both $\cos^2\psi\bkO$ of the intruder orbits with the positive  parity at
$\omega=0$[MeV/$\hbar$] in the proton and neutron fields,
\begin{equation}
\label{psi1}
    \cos^2\psi\bkO \approx \sum_{n_x+n_y+n_z=even}
    |\lang n_x,n_y,n_z ; \sigma |\mu\bkO\rang|^2,
\end{equation}
have maximal values at
$\omega\sim 1.3$[MeV/$\hbar$] and  $\omega\sim 1.35 $[MeV/$\hbar$],
where the components of the IB moment of inertia,
${\cal J}_{p,s=+1}\bkO$ and ${\cal J}_{n,s=+1}\bkO$,
show maxima, respectively (Fig. \ref{fig:psi} (b)).

The dynamical and IB moments of
inertia also increase at $\omega\sim .9$ [MeV/$\hbar$]
and show maxima around $\omega\sim 1.2$ [MeV/$\hbar$]
(Fig.\ref{fig:mois}(b)).
However, the behavior of
the $\moi\bkO$ and $\moiing\bkO$
moment of inertia  is less correlated
at larger values of the rotational frequency.
The pseudo-crossing is much sharper in $^{56}Ni$
in comparison with the one in $^{32}S$ (Fig.\ref{fig:sprs}),
i.e. the octupole interaction is expected to be weaker.
Both $\cos^2\psi\bkO$ decrease rapidly at $\omega > 1.55 $[MeV/$\hbar$]
(see Fig.(\ref{fig:psi}(b)).
Due to  the termination of the  parity mixing of the
s.p. levels, the matrix element
$J_{ij}\bkO$  between mixed states, Eq.(\ref{jij}), 
becomes rapidly small.
As a result, the IB moment of inertia,  $\moiing\bkO$, 
also decreases. On the other hand, the
dynamical moment of inertia, $\moi\bkO$, reflects changes
of the self-consistent mean field  and, in particular, 
the modification of the two-body interaction due 
to the rotation. Due to the two-body interaction,
the high intruder unoccupied state ($s=+$)
contributes to the sudden increase of the dynamical moment of
inertia  observed
at  high rotational frequencies  $\omega >1.5[$MeV$/\hbar]$.

The fluctuations of octupole deformations
$\Delta Q_{31}$ and $\Delta Q_{33}$ are always larger than 1 in
all region of the ND band (Fig. \ref{fig:fluc}(b)),
in spite of the presence of the octupole minimum (Fig. \ref{fig:pesS}(b)).
The octupole minimum  in $^{56}Ni$ is much more
shallow in comparison with  the one of $^{32}S$ and it could explain
large fluctuations of the octupole deformations in the
considered case.

\section{Summary}

It was expected from various calculations based on the MMM
that octupole deformations
would arise in rotating nuclei \cite{Ab}.
Using the cranking HF(B) approach with the effective Gogny
interaction, we found that the non-axial octupole deformation
associated with the $Y_{31}$ term in the octupole family becomes
important in the yrast band of $^{32}S$
at the angular momenta $I\geq5[\hbar]$.
The primary mechanism behind the occurrence of the octupole
deformation is related
to the strong mixing via octupole interaction
of s.p. orbits with a positive  simplex quantum number.
Similar phenomena  have been observed in the nucleus $^{56}Ni$,
where we predict the octupole softness
in the excited ND band at high spins $I\geq10[\hbar]$.

Finally, an exploration of the octupole phenomenon certainly could
deepen our understanding of different aspects of 
a spontaneous symmetry breaking mechanism in finite Fermi 
systems like nuclei. In particular, the breaking of 
the intrinsic reflection symmetry could be related to
unexpected strong electric dipole and octupole 
transitons in rotational bands and 
to a formation of multi-cluster structures in 
strongly deformed nuclei.
Measurements using new generations of modern detectors
can test the predictions made within
the  HF(B) approach and lead to new
insights regarding effective nucleon-nucleon
interactions and their properties.

\section*{Acknowledgements}

We are thankful to F. Sakata, Y. Hashimoto
and Y. Kanada-En'yo for their support in numerical
calculations. We are also grateful to K. Matsuyanagi
for illuminating communications.
This work was supported in part by the RFFI under
Grant 00-02-17194.
\appendix
\def\theequation{\thesection.\arabic{equation}}
\section{A two-level model}
\setcounter{equation}{0}
\label{sec:octth}

Let us consider the situation such that the high-lying intruder s.p. orbit
$\mu$ is coming down to the highest-lying occupied orbit $\nu$
due to the Coriolis term $-\omega\hjz$;
they have the same simplex quantum number.
For the sake of simplicity, we restrict our discussion
within the $2\times2$ subspace.
We assume that
i) our states are
eigenstates of the parity operator $P$ at $\omega=0$;
ii) there are no changes in the subspace
expanded by the simplex partners of  $\mu $ and $\nu$
at $\omega \neq 0$.

At $0\leq \omega\leq \omega_c$ the eigenstates
can be written as
\begin{eqnarray}
\label{spwf}
  |      \nu  \bkO \rang &=& \cos \psi\bkO
  |      \nu  \bkOz\rang  +  \sin \psi\bkO
  |     \mu \bkOz\rang,  \nonumber\\
  |     \mu \bkO \rang &=& \cos \psi\bkO
  |     \mu \bkOz\rang  -  \sin \psi\bkO
  |      \nu \bkOz\rang,
\end{eqnarray}
where $|\psi\bkO|$ is a monotonic
function of $\omega$
\footnote{A sign of $\psi\bkO $ is not important
for the absolute value of the octupole deformation.},
which satisfies the following conditions;
\begin{eqnarray}
 &&0\leq|\psi\bkO |\leq \frac{\pi}{2}, \\
 &&\psi\bkOz=0, \quad  |\psi(\omega_c)|=\frac{\pi}{2}. \nonumber
\end{eqnarray}
Here, $\omega_c$ is the largest rotational frequency.
Using the unitary transformation in Eq.(\ref{spwf}),
at $\omega \neq 0$ the density matrix has the following
form in this subspace
\begin{eqnarray}
\label{ttrho}
     \mr\bkO &\equiv&
\left(
\begin{array}{cc}
\rho_{\nu \nu}\bkO & \rho_{\nu \mu}\bkO \\
\rho_{\mu \nu}\bkO & \rho_{\mu\mu}\bkO
\end{array}
\right)\nonumber\\ &=&
\left(
\begin{array}{cc}
 \cos  \psi\bkO    &  \sin  \psi\bkO   \\
-\sin  \psi\bkO    &  \cos  \psi\bkO
\end{array}
\right)                    \cdot
\left(
\begin{array}{cc}
      1             &       0            \\
      0             &       0
\end{array}
\right)\nonumber\\&& \quad \cdot
\left(
\begin{array}{cc}
 \cos  \psi\bkO    & -\sin  \psi\bkO   \\
 \sin  \psi\bkO    &  \cos  \psi\bkO   \\
\end{array}
\right)         \\ &=&
\left(
\begin{array}{cc}
 \cos^2\psi\bkO    & -\sin 2\psi\bkO /2\\
-\sin 2\psi\bkO /2 &  \sin^2\psi\bkO
\end{array}
\right).\nonumber
\end{eqnarray}

At $\omega=0$,
the matrix elements of the octupole operator $\hq_3$
have the following structure
\begin{equation}
  \label{ttq}
     \mq   \equiv
\left(
\begin{array}{cc}
q_{\nu \nu} & q_{\nu \mu} \\
q_{\mu \nu} & q_{\mu\mu}
\end{array}
\right)=
\left(
\begin{array}{cc}
0 & q \\
q & 0
\end{array}
\right),
\end{equation}
where $q$ is a finite real number. At $\omega \neq 0$
the expectation value of $\hq_3$ is
\begin{equation}
\label{expq}
|\lang \hq_3 \rang|
=|{\rm tr}(     \mq       \mr\bkO )|
= |q \sin 2  \psi\bkO |, \nonumber
\end{equation}
due to the assumption ii).

At $\omega=0$ matrix elements of the angular momentum operator $\hjz$
are defined as
\begin{equation}
\label{ttj}
\mj \equiv
\left(
\begin{array}{cc}
j_{\nu \nu} & j_{\nu \mu} \\
j_{\mu \nu} & j_{\mu\mu}
\end{array}
\right) =
\left(
\begin{array}{cc}
j & 0  \\
0 & j'
\end{array}
\right).
\end{equation}
At $\omega \neq 0$, using the unitary transformation
in Eq.(\ref{spwf}), we obtain for the matrix elements of $\mj$
\begin{eqnarray}
\label{jth}
&&
\left(
\begin{array}{cc}
j_{\nu \nu}\bkO & j_{\nu \mu}\bkO \\
j_{\mu \nu}\bkO & j_{\mu\mu}\bkO
\end{array}
\right)=\\
&&
\left(
\begin{array}{cc}
    j \cos^2\psi\bkO  + j'\sin^2\psi\bkO  &
(j'-j)\sin 2\psi\bkO          /2          \\
(j'-j)\sin 2\psi\bkO          /2           &
    j'\cos^2\psi\bkO  + j \sin^2\psi\bkO
\end{array}
\right), \nonumber
\end{eqnarray}
From Eqs. (\ref{expq}) and (\ref{jth}),
it follows that
$|\lang\hq_3\rang|$ and
$|j_{\mu \nu}\bkO|^2=(j'-j)^2\sin^2 2\psi\bkO /4$ have a maximum at
$|\psi\bkO |=\pi/4$,
where there is a strong mixing of the unperturbed eigenfunctions.
Therefore, the $\mu \nu$ component of the Inglis-Beliaev
moment of inertia
\begin{eqnarray}
  \label{jmum}
\frac{2|j_{\mu \nu}\bkO|^2}
{\epsilon_\mu\bkO + \epsilon_\nu\bkO}=
\frac{(j'-j)^2\sin^2 2\psi\bkO}
{2(\epsilon_\mu\bkO + \epsilon_\nu\bkO)}
\end{eqnarray}
can be large enough
to affect the total value of the moment of inertia
due to the smallness of its denominator
$\epsilon_\mu\bkO + \epsilon_\nu\bkO$ at
this rotational frequency.


\begin{center}
{\bf  Figure Captions}
\end{center}
\begin{figure}[htbp]
  \begin{center}
    \leavevmode
    \caption[Total binding energies vs. $I$.]
{The total binding energy in $^{32}S$ as a function of the angular
momentum.
The calculated points indicated by symbols `$+$' and `$\times$' 
for ND and SD states, respectively, are connected by solid lines.
Experimental values for states with different parities and spins 
are indicated by symbols: filled square is used 
 for the positive parity  and even values of $I$;
open square is used for the negative parity  and even values of $I$;
filled triangle is used for the positive parity and  odd values of $I$;
inverse open triangle is used for the negative parity and 
odd values of $I$. The experimental values of the total binding energy and
excited levels  from Refs. \cite{Aud93} and \cite{NDSS}, respectively.}
    \label{fig:enes}
  \end{center}
\end{figure}
\begin{figure}[htbp]
  \begin{center}
    \leavevmode
    \caption[$I$-$\omega$:$^{32}S$]{The angular momentum $I$ versus
     the rotational frequency $\omega $
    in $^{32}S$ and $^{56}Ni$.
    The calculated values of the angular momentum connected by 
    solid lines are indicated by symbols: 
    filled square and filled circle are  used for ND and SD states,
    respectively, for $^{32}S$; open square and open circle are
    used for ND and SD states, respectively, for $^{56}Ni$; 
    `$+$' and `$\times$' are used for the results of calculations with
    the Skyrme III interaction \cite{Yam98}.}
    \label{fig:oms}
  \end{center}
\end{figure}
\begin{figure}[htbp]
  \begin{center}
    \leavevmode
   \caption[$\beta$-deformations.]
    {$\beta$-deformations in $^{32}S$ and $^{56}Ni$ as 
    a function of the rotational frequency $\omega$.
     The symbols are used here are defined in 
     Fig.\ref{fig:oms}}
     \label{fig:betas}
  \end{center}
\end{figure}
\begin{figure}[htbp]
  \begin{center}
    \leavevmode
    \caption[The octupole deformations in $^{32}S$
             and $^{56}Ni$.]
            {The non-axial octupole deformation: (a) in the 
             ND yrast band of $^{32}S$
             and (b) in the ND excited band of  $^{56}Ni$, -- 
             as a function of the rotational frequency $\omega$.
     The calculated values connected by solid lines are indicated
      by symbols: filled square is used for
      $|\beta_{31}|$; open square is used  for $|\beta_{33}|$.
     SIII means the results of Ref. \cite{Yam98}.}
    \label{fig:octs}
  \end{center}
\end{figure}
\begin{figure}[htbp]
  \begin{center}
    \leavevmode
    \caption[Single-particle energies in $^{32}S$ and $^{56}Ni$.]
{Neutron single-particle energies near the Fermi surface versus
the rotational frequency $\omega$ : 
(a) in $^{32}S$ and
(b) in  $^{56}Ni$. 
The occupied states with different simplex quantum number $s$
connected by lines are indicated by symbols:
filled square and open square are used for 
the $s=+1$ and $s=-1$ states, respectively. 
The unoccupied states with  the simplex  $s=+1$ and $s=-1$
are connected by solid and dotted lines, 
respectively.}
    \label{fig:sprs}
  \end{center}
\end{figure}
\begin{figure}[htbp]
  \begin{center}
    \leavevmode
    \caption[Density distributions in $^{32}S$ and $^{56}Ni$.]
{The density distribution:
(a) in the ND yrast   band of $^{32}S$  at $\omega=2.2$ [MeV/$\hbar$]; 
(b) in the ND excited band of $^{56}Ni$ at $\omega=1.4$ [MeV/$\hbar$].}
    \label{fig:dens}
  \end{center}
\end{figure}
\begin{figure}[htbp]
  \begin{center}
    \leavevmode
    \caption[Moments of inertia in $^{32}S$ and $^{56}Ni$.]
    {The  static, $\moist\bkO$, the dynamical, $\moi\bkO$, and 
     the Inglis-Beliaev, $\moiing\bkO$, moments of inertia
     versus the rotational frequency $\omega$ :
     (a) in the ND yrast   band in $^{32}S$;
     (b) in the ND excited band in $^{56}Ni$. 
      See the definition of  different moments of inertia in the text.}
    \label{fig:mois}
  \end{center}
\end{figure}
\begin{figure}[htbp]
  \begin{center}
    \leavevmode
    \caption[Components of $\moiing\bkO$ in $^{32}S$.]
       {Components of the Inglis-Beliaev, $\moiing\bkO$, moment of
        inertia as a function of the 
       rotational frequency $\omega$ : 
        (a) in $^{32}S$ and (b) in $^{56}Ni$.
        Notations of components are defined in Eq. (\ref{jingp1}). }
    \label{fig:jinp}
  \end{center}
\end{figure}
\begin{figure}[htbp]
  \begin{center}
    \leavevmode
    \caption[Degree of parity mixing of intruder orbit.]
       {The degree of mixing of intruder orbits, $\cos^2\psi\bkO$,
        in the region of quasi-crossing:
        (a) in $^{32}S$ and (b) in $^{56}Ni$. 
        The quantity  $\cos^2\psi\bkO$ is defined in the text. }
    \label{fig:psi}
  \end{center}
\end{figure}
\begin{figure}[htbp]
  \begin{center}
    \leavevmode
    \caption[Fluctuation of oct. and quad. moments.]
    {The inverse of fluctuation of the octupole and quadrupole moments 
     as a function of the rotational frequency $\omega$ :
         (a) in the ND yrast band   in $^{32}S$; 
         (b) in the ND excited band in $^{56}Ni$. The fluctuation 
       $\Delta Q_{lm}$ is defined by Eq.(\ref{fluc1}).}
    \label{fig:fluc}
   \end{center}
\end{figure}
\begin{figure}[htbp]
  \begin{center}
    \leavevmode
    \caption[Potential energy surface along the $\beta_{31}$
             direction.]
       {The potential energy curve along the $\beta_{31}$
             direction: (a) in $^{32}S$ and (b) in $^{56}Ni$. 
        The calculated values of the $\beta_{31}$ indicated by 
        symbols are connected by solid line.}
    \label{fig:pesS}
  \end{center}
\end{figure}
\begin{figure}[htbp]
  \begin{center}
    \leavevmode
    \caption[Total binding energies and charge radiuses
             in nickel isotopes.]
       {(a)The total binding energy and (b) the charge radius
             in nickel isotopes.\\
Experimental values of total binding energies from
Ref. \cite{Aud93}.
        A charge radius $r_c$ is calculated with the approximation
        $r_c^2 = r_p^2 + .64$ [fm$^2$] \cite{Dob96} from the
        calculated values of proton radius $r_p$.
Experimental values of $r_c$ for $^{58,60}$Ni
from Ref. \cite{Nad96};
for $^{56}Ni$ it is the extrapolated value from
Ref. \cite{Nic92}.}
    \label{fig:nick}
  \end{center}
\end{figure}
\begin{figure}[htbp]
  \begin{center}
    \leavevmode
    \caption[Total binding energy vs. $I$.]
{The total binding energy in $^{56}Ni$ as a function of the
angular momentum.
The calculated points indicated by symbols `$+$' and `$\times$' 
for SD and ND states, respectively, are connected by solid lines.
Experimental values for different states  
are indicated by symbols: filled square is used 
 for yrast non-rotating states;
open square is used for the excited states whose binding
energies are proportional to $I(I+1)$ [$\hbar^2$] starting from
the $0^+_3$ state; filled circle is used for other states.
The calculations of the ground spherical state reproduce
the experimental  ground binding energy quite well. 
Experimental values of the total binding energy and  excited levels
from Refs. \cite{Aud93} and \cite{NDSNi}, respectively.}
    \label{fig:eneni}
  \end{center}
\end{figure}
\begin{figure}[htbp]
  \begin{center}
    \leavevmode
    \caption[Single-particle energies in the SD band of $^{56}Ni$.]
    {Neutron single-particle energies of $^{56}Ni$
    along the SD band near the Fermi surface. 
    The symbols are used here are defined in Fig.\ref{fig:sprs}.}
    \label{fig:sprnisd}
  \end{center}
\end{figure}
\begin{figure}[htbp]
  \begin{center}
    \leavevmode
    \caption[Density distribution of SD minimum at $\omega=0$ in $^{56}Ni$.]
       {The density distribution of SD minimum at
        $\omega=0$ [MeV/$\hbar$] in $^{56}Ni$.}
    \label{fig:sddni}
  \end{center}
\end{figure}

\begin{thebibliography}{99}
\bibitem{BM75}     A. Bohr and B. Mottelson, {\it Nuclear Structure},
                   Vol.2 (Benjamin, New York, 1975)
\bibitem{RS}       P. Ring and P.Schuck, {\it The Nuclear Many--Body
                   Problem} (Springer--Verlag, Berlin, 1980)
\bibitem{Ab}       S. \AA berg, H. Flocard and W. Nazarewicz,
                   Annu. Rev. Nucl. Part. Sci. {\bf 40}, 439 (1990).
\bibitem{col}      Z. Patyk, A. Baran, J.F. Berger, J. Decharg{\'e},
                   J. Dobaczewski, P. Ring and A. Sobiczewski,
                   Phys.Rev. {\bf C59}, 704 (1999).
\bibitem{Nol}      P.J. Nolan and P.J. Twin, Annu. Rev. Nucl.Part.
                   Sci. {\bf 38}, 533 (1988).
\bibitem{Jan}      R.V.F. Janssens and T.L. Khoo, Annu. Rev. Nucl.Part.
                   Sci. {\bf 41}, 321 (1991).
\bibitem{Bak}      C. Baktash, B. Haas and W. Nazarewicz,
                   Annu. Rev. Nucl.Part. Sci. {\bf 45}, 485 (1995).
\bibitem{Bak1}     C. Baktash, Prog. Part. Nucl. Phys.
                   {\bf 38}, 291 (1997).
\bibitem{But}      P.A. Butler and W. Nazarewicz,
                   Rev. Mod. Phys. {\bf 68}, 350 (1996).
\bibitem{Frau}     S. Frauendorf,
                   submitted to Rev. Mod. Phys.
\bibitem{Nak}      T.Nakatsukasa, K.Matsuyanagi and S.Mizutori,
                   Progr. Theor. Phys. {\bf 87}, 607 (1992).
\bibitem{RN}       R. Nazmitdinov and S. \AA berg,
                   Phys.Lett. {\bf B289}, 238 (1992).
\bibitem{Ch}       R.R. Chasman, Phys.Lett. {\bf B266}, 243 (1991).
\bibitem{Dud}      X. Li and J. Dudek, Phys.Rev. {\bf C49}, R1250 (1994).
\bibitem{Tak98}    S. Takami, K. Yabana and M. Matsuo,
                   Phys. Lett. {\bf B341}, 242 (1998).
\bibitem{Yam98}    M. Yamagami and K. Matsuyanagi,
                   {Proc. Int. Conf. on Nuclear Structure '98},
                   edited by C. Baktash
                   (Gatlinburg, Tennessee, 1998), p.327;
                   nucl-th/9908060.
\bibitem{prl}      W. D. Heiss, R. G. Nazmitdinov and S. Radu,
                   Phys. Rev. Lett. {\bf 72}, 2351 (1994);
                   Phys. Rev. {\bf B51}, 1874 (1994);
                   Phys. Rev. {\bf C52}, 3032 (1995).
\bibitem{mat}      A. Arita and K. Matsuyanagi, 
                   Prog. Theor. Phys. {\bf 91}, 723 (1994);
                   Nucl. Phys. {\bf A592}, 9 (1995).
\bibitem{He98}     W. D. Heiss, R. A. Lynch and R. G. Nazmitdinov,
                   JETP Lett.   {\bf 69}, 563 (1998).
\bibitem{He99}     W. D. Heiss, R. A. Lynch and R. G. Nazmitdinov,
                   Phys. Rev. {\bf C60}, 034303 (1999).
\bibitem{Gog75}    D. Gogny, {\it Nuclear Self-consistent Fields},
                   edited by G. Ripka and M. Porneuf,
                   (North-Holland, Amsterdam, 1973) p.333;
\bibitem{Dev80}    J. Decharg{\'e} and D. Gogny, Phys. Rev.
                   {\bf C21}, 1568 (1980).
\bibitem{Gir83}    M. Girod and B. Grammaticos,
                   Phys. Rev. {\bf C27}, 2317 (1983).
\bibitem{Ber84}    J. F. Berger, M. Girod and D. Gogny,
                   Nucl. Phys. {\bf A428}, 23c (1984);
                   Nucl. Phys. {\bf A502}, 85c (1989);
                   Comp. Phys. Comm. {\bf 63}, 365 (1991).
\bibitem{Egi93}    J. L. Egido and L. M. Robledo, Phys. Rev. Lett.
                   {\bf 70}, 2876 (1993);
                   J. L. Egido,    L. M. Robledo and R. R. Chasman
                   Phys. Lett. {\bf B322}, 22 (1994).
\bibitem{Gir94}    M. Girod, J. P. Delaroche, J. F. Berger and J. Libert,
                   Phys. Lett. {\bf B325}, 1 (1994).
\bibitem{Cau95}    E. Caurier, J. L. Egido, G. Mart{\'\i}nez-Pinedo,
                   A. Poves, J. Retamosa, L. M. Robledo
                   and A. P. Zuker,
                   Phys. Rev. Lett. {\bf 75} 2466 (1995);
                   G. Mart{\'\i}nez-Pinedo, A. Poves, L. M. Robledo,
                   E. Caurier, F. Nowacki, J. Retamosa, and A. P. Zuker,
                   Phys. Rev. {\bf C54}, R2154 (1996);
                   S. M. Lenzi {\it et al.},
                   Phys. Rev. {\bf C56}, 1313 (1997).
\bibitem{Tan98}    T. Tanaka, F. Sakata, and K. Iwasawa,
                   Phys. Rev. {\bf C58}, 2765 (1998).
\bibitem{Tan97}    T. Tanaka, F. Sakata, T. Marumori and K. Iwasawa,
                   Phys. Rev. {\bf C56}, 180 (1997).
\bibitem{Iwa97}    K. Iwasawa, F. Sakata, T. Tanaka,
                   Y. Hashimoto and T. Marumori,
                   {\it Progress in Particle and Nuclear Physics}
                   {\bf 38}, 249 (Elsevier Science, 1997).
\bibitem{Dob99}    J. Dobaczewski, J. Dudek, S. G. Rohozinski
                   and T. R. Werner,
                   nucl-th/9912072; nucl-th/9912073.
\bibitem{Flo73}    H. Flocard, P. Quentin, A. K. Kerman
                   and D. Vautherin,
                   Nucl. Phys. {\bf A203}, 433 (1973).
\bibitem{Goo79}    A. L. Goodman, in {Advances in Nuclear Physics},
                   edited by J. W. Negele and E. Vogt
                   (Plenum, New York, 1979), Vol. 11, p. 293.
\bibitem{TV}       D.J. Thouless and J.G. Valatin,
                   Nucl.Phys. {\bf 31}, 211 (1962).
\bibitem{Aud93}    G. Audi and A. H. Wapstra, Nucl. Phys.
                   {\bf A565}, 1 (1993).
\bibitem{NDSS}     P. M. Endt, Nucl. Phys. {\bf A521}, 1 (1990).
\bibitem{Mol99}    H. Molique, J. Dobaczewski and J. Dudek,
                   nucl-th/9907103.
\bibitem{KN}       J. Kvasil and R.G. Nazmitdinov,
                   Nucl. Phys. {\bf A439}, 86 (1985)
\bibitem{DAN}      F.~D\"onau, D. Almehed and R.G. Nazmitdinov,
                   Phys.Rev.Lett. {\bf 83}, 280 (1999).
\bibitem{Dob96}    J. Dobaczewski, W. Nazarewicz and T. R. Werner,
                   Z. Phys. {\bf A354} 27 (1996).
\bibitem{Nad96}    E. G. Nadjakov, K. P. Marinova and Yu. P. Gangrsky,
                   At. Nucl. D. Tab. , {\bf 56}, 133 (1994).
\bibitem{Nic92}    B. A. Nikolaus T. Hoch and D. G. Madland,
                   Phys. Rev. {\bf C46}, 1757 (1992).
\bibitem{NDSNi}    H. Junde, Nuclear Data Sheets, {\bf 67}, 523 (1992).
\bibitem{MN}       E. R. Marshalek and R. G. Nazmitdinov,
                   Phys. Lett. {\bf B300}, 199 (1993);
                   E. R. Marshalek, R. G. Nazmitdinov and I. Ragnarsson,
                   Bulletin of Russian Academy of Sciences: Physics,
                   {\bf 57}, 1709 (1993).
\bibitem{Sve97}    C. E. Svensson {\it et al.},
                   Phys. Rev. Lett. {\bf 79} 1233 (1997);
                   Phys. Rev. Lett. {\bf 82} 3400 (1999);
                   M. Devlin {\it et al.},
                   Phys. Rev. Lett. {\bf 82}, 1233 (1999).
\bibitem{She72}    R. K. Sheline, I. Ragnarsson and S. G. Nilsson,
                   Phys. Lett. {\bf B41}, 115 (1972).
\bibitem{Ben84}    T. Bengtsson, M. Faber, M. Ploszajczak,
                   I. Ragnarsson and S. {\AA}berg,
                   Lund-MPh-84/01.
\bibitem{ND}       W. Nazarewicz and J. Dobaczewski,
                   Phys. Rev. Lett. {\bf 68}, 154 (1992).
\bibitem{Bet81}    R. R. Betts et al.,
                   Phys. Rev. Lett. {\bf 47}, 23 (1981).
                   R. R. Betts,
                   Nucl. Phys. {\bf A447}, 257c (1985).
\bibitem{Ueg94}    E. Uegaki and Y. Abe,
                   Phys. Lett. {\bf B340}, 143 (1994).
\bibitem{Nou99}    R. Nouicer et al.,
                   Phys. Rev. {\bf C60},  041303(1999).
\end{thebibliography}
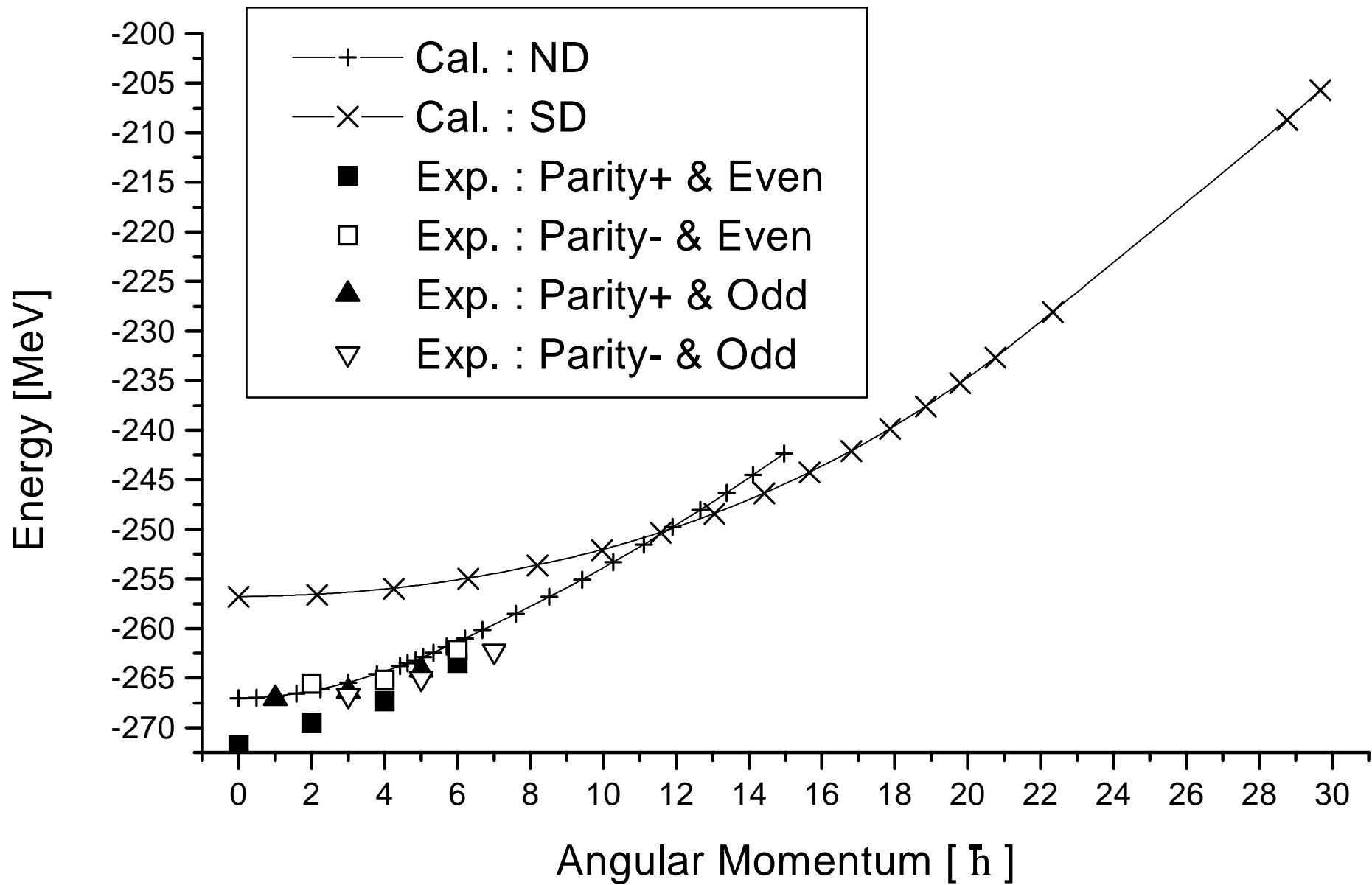
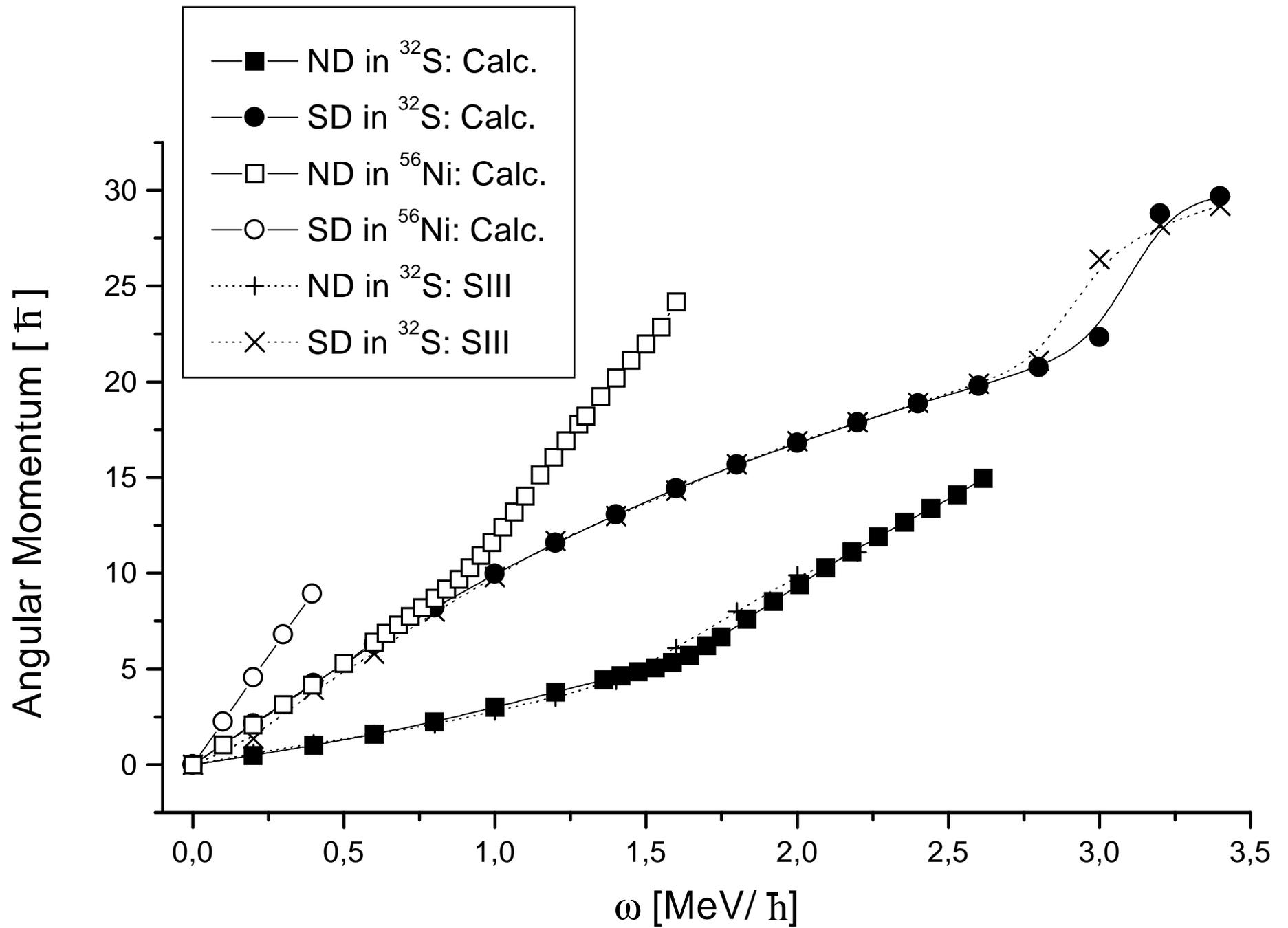
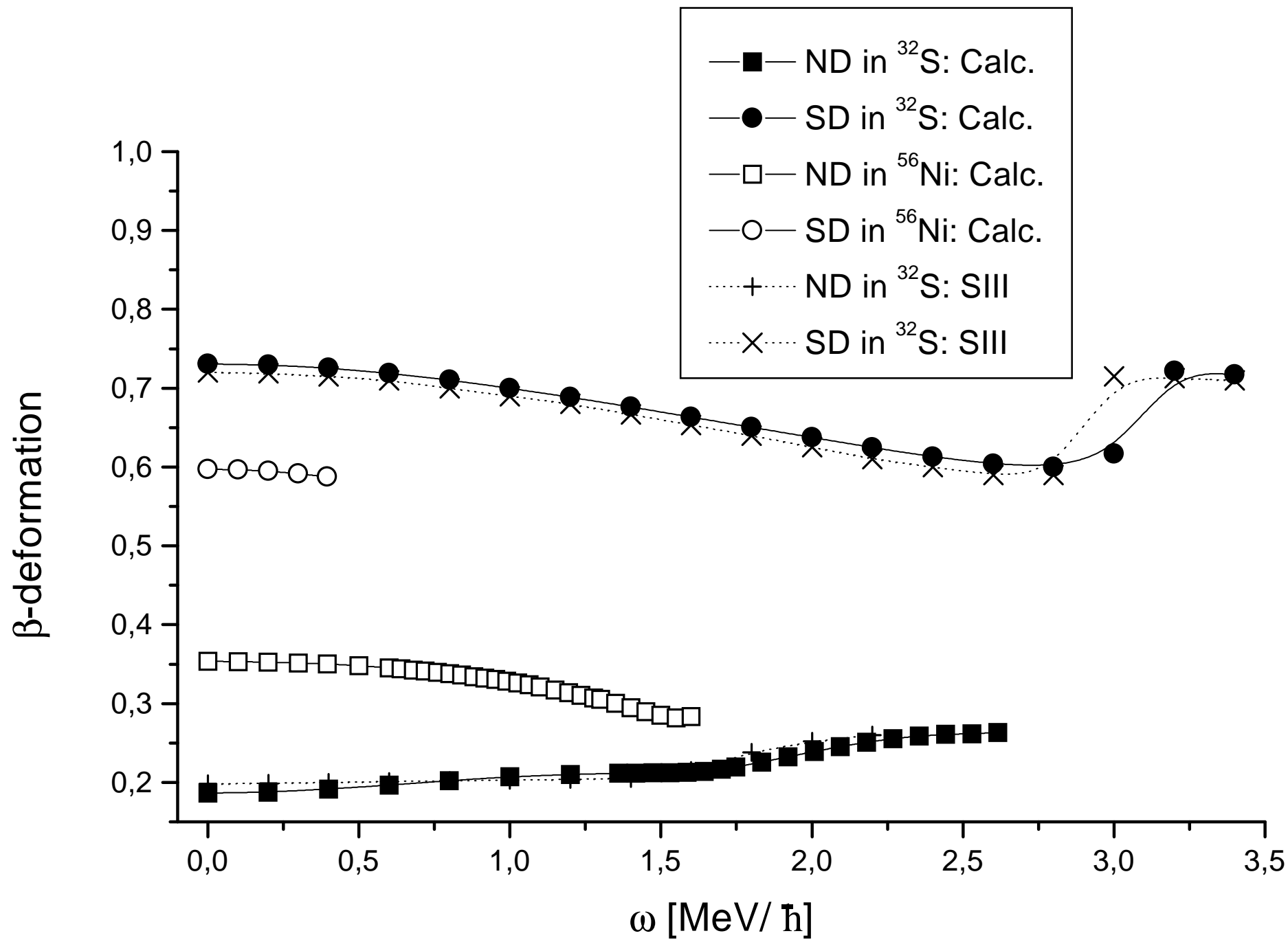
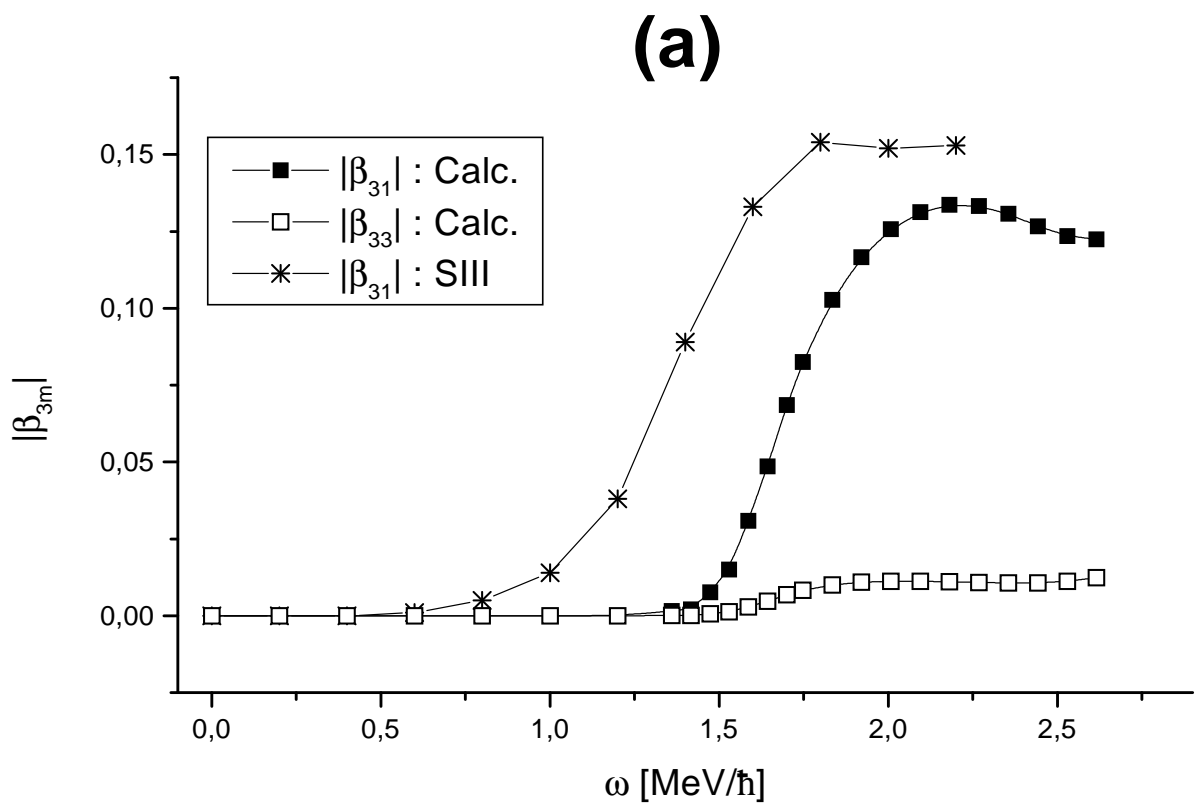
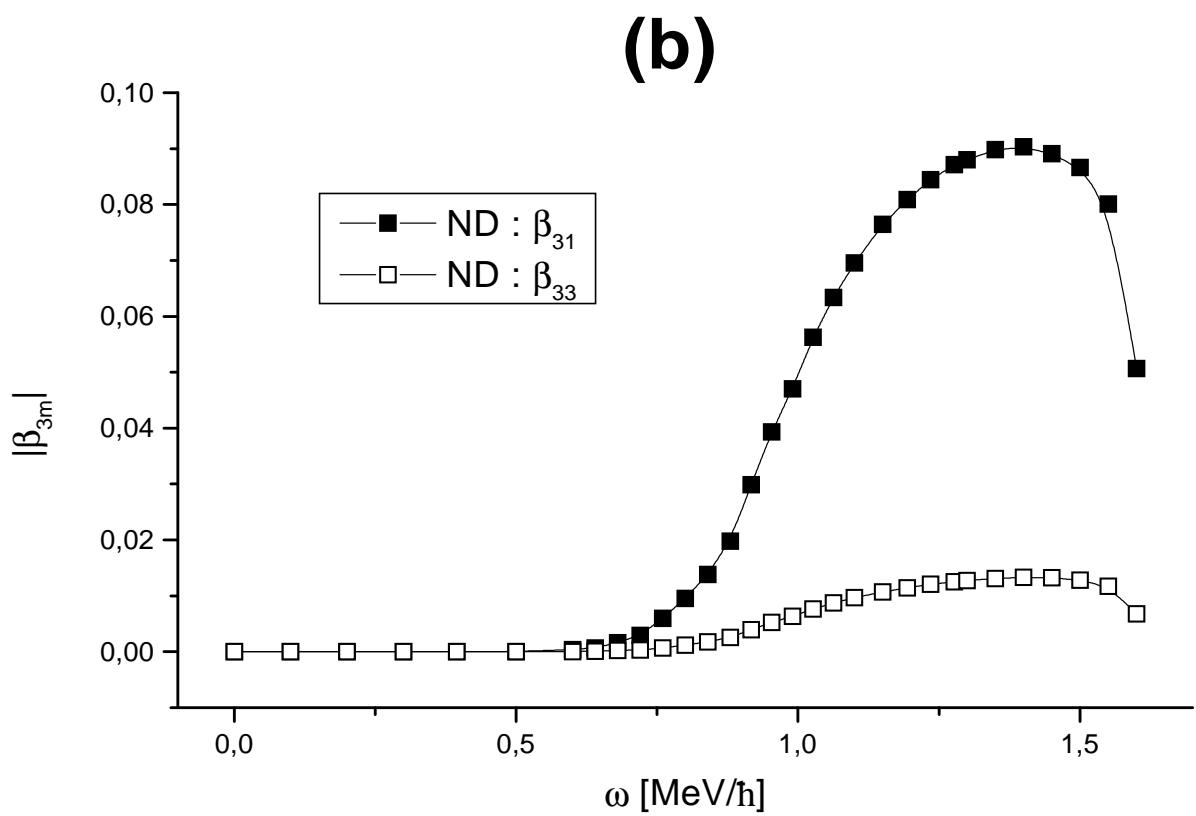
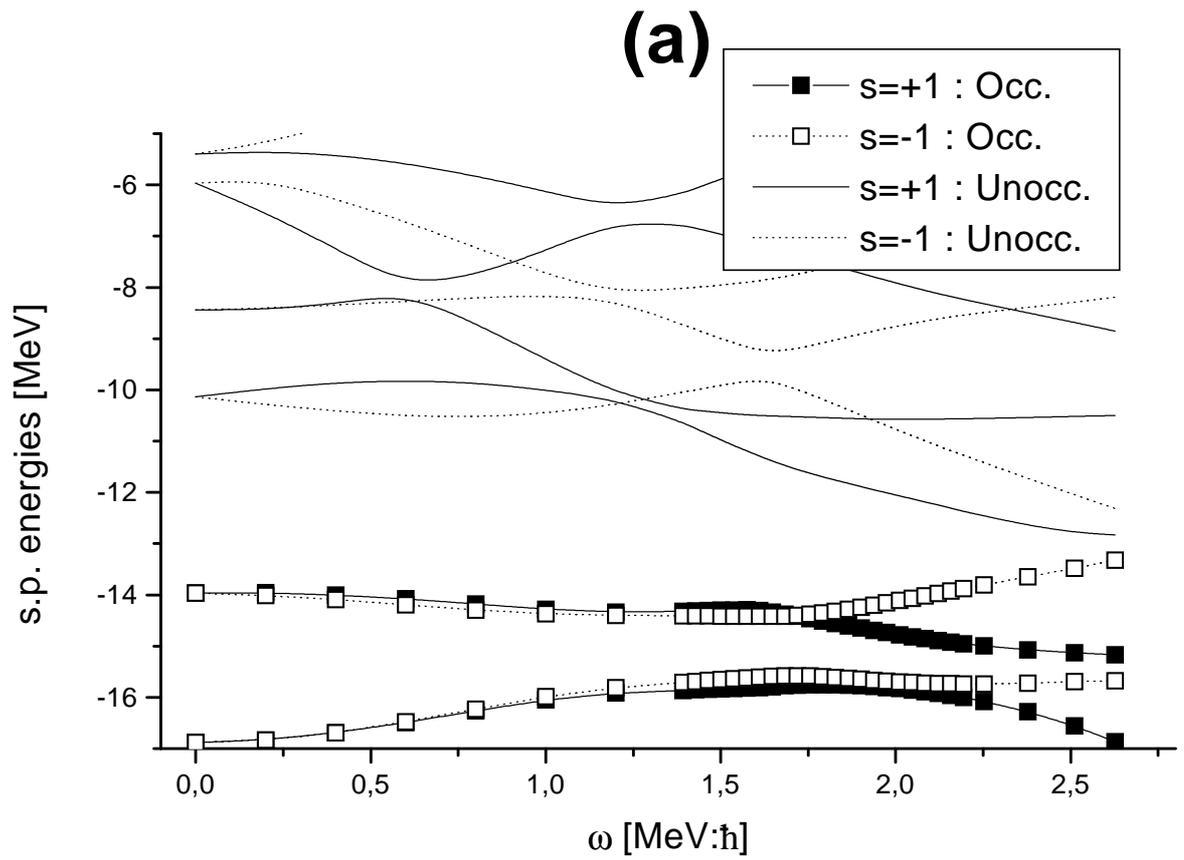
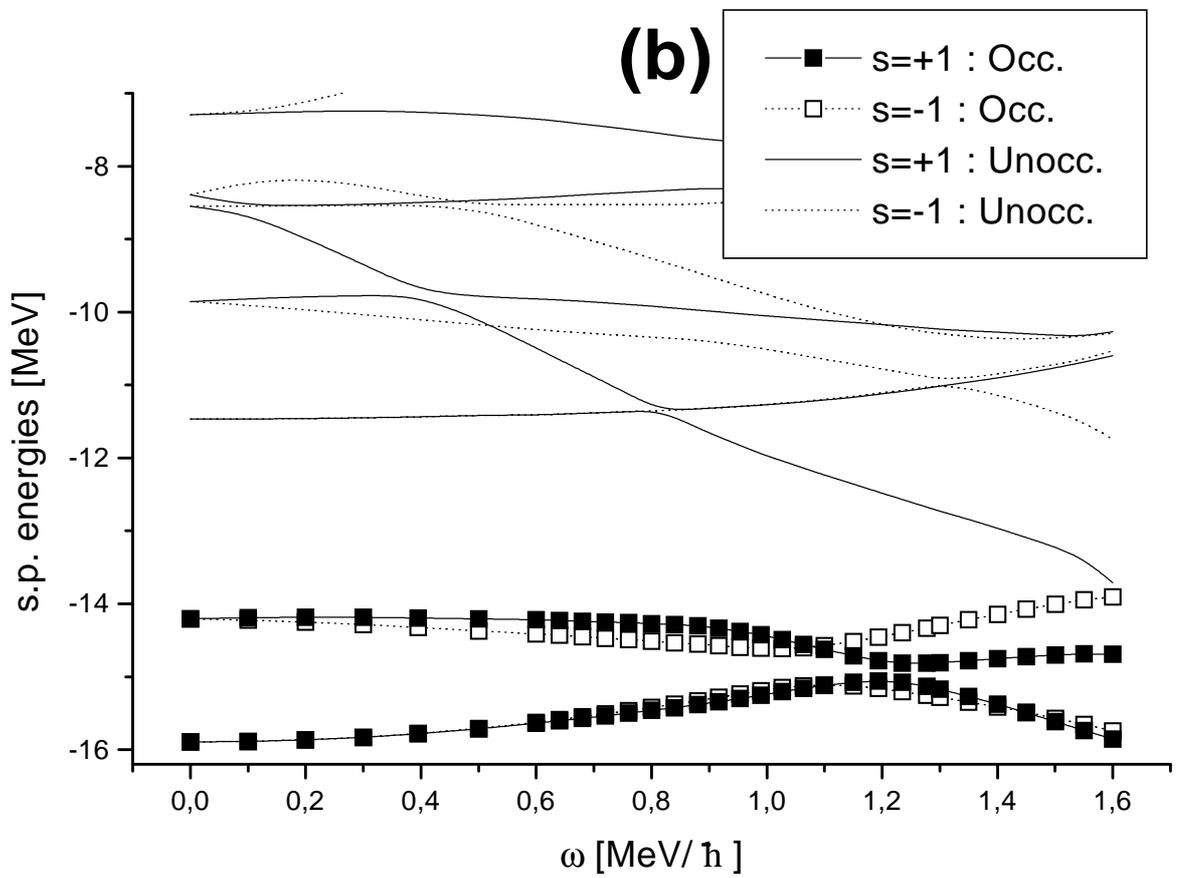
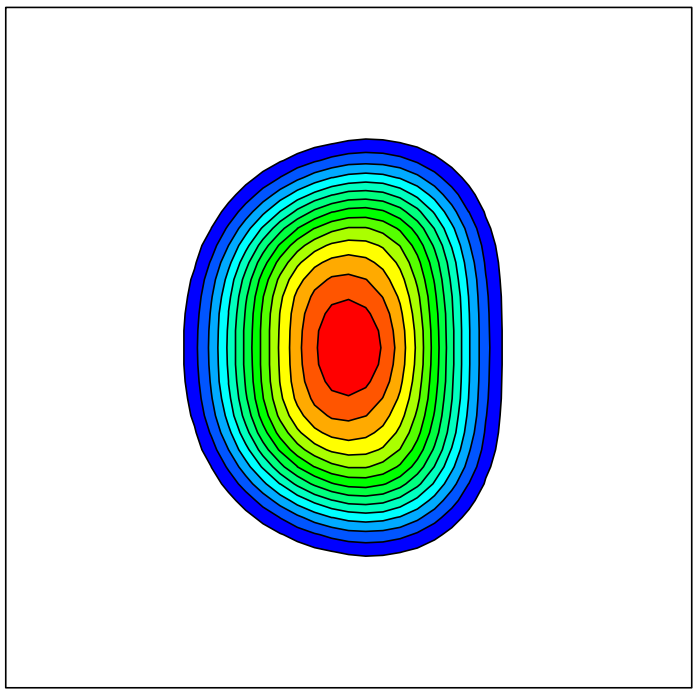
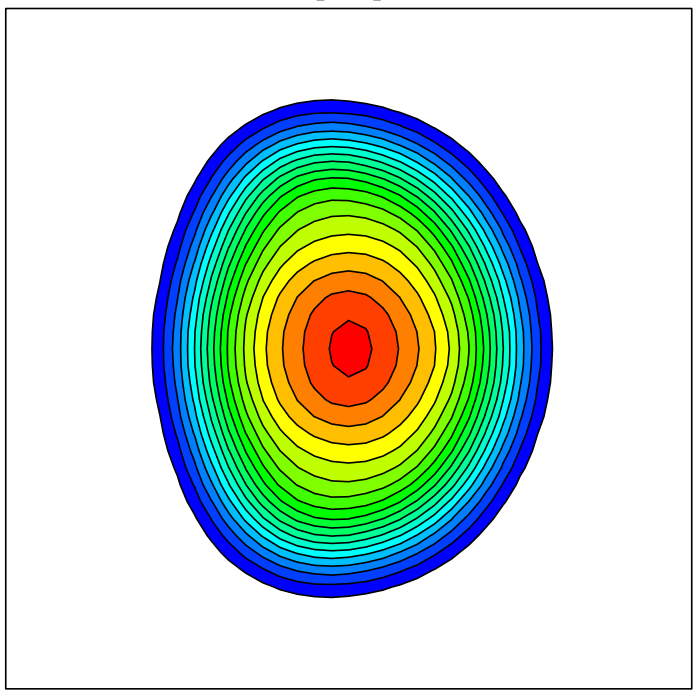
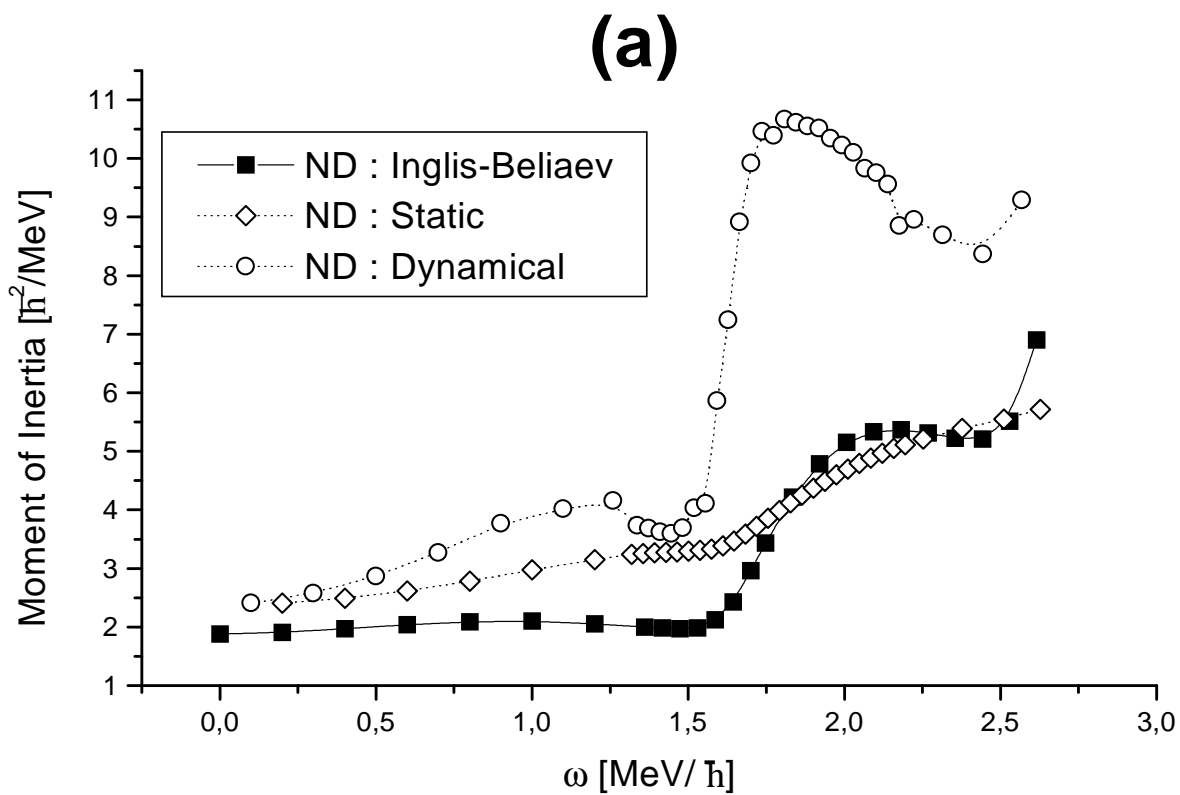
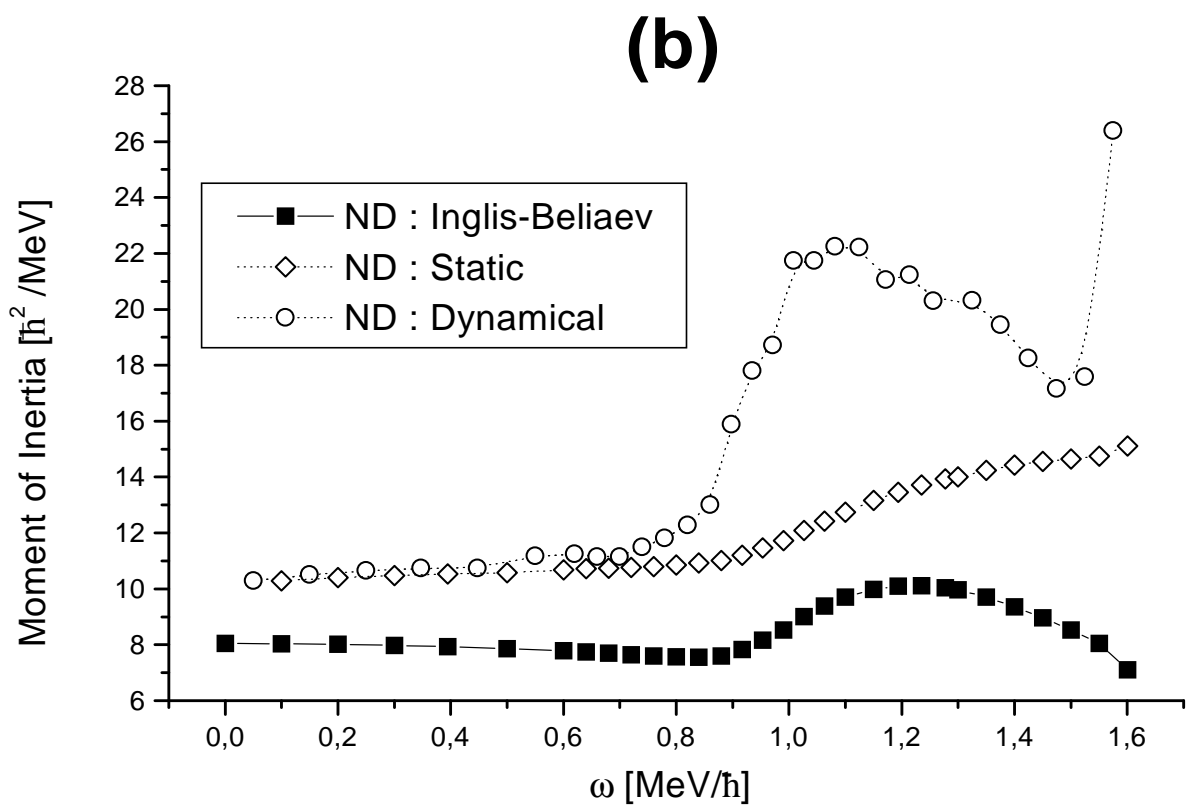
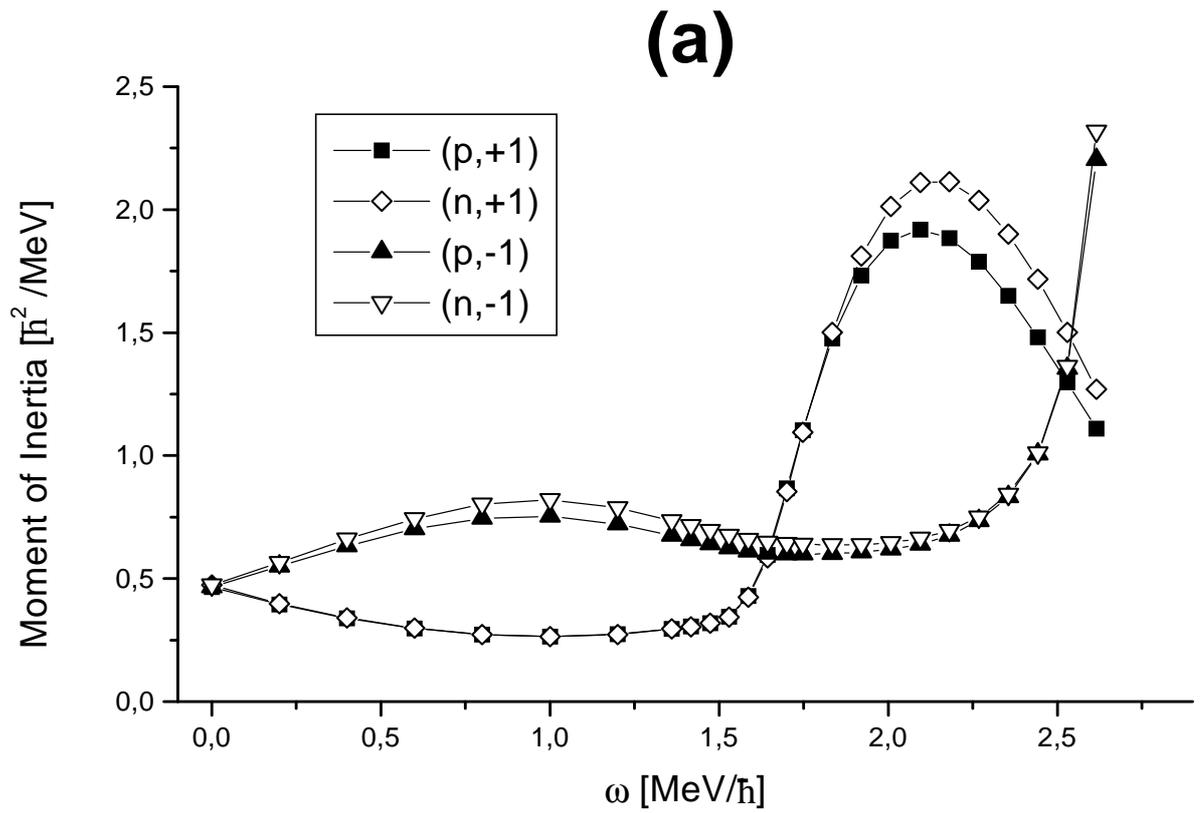
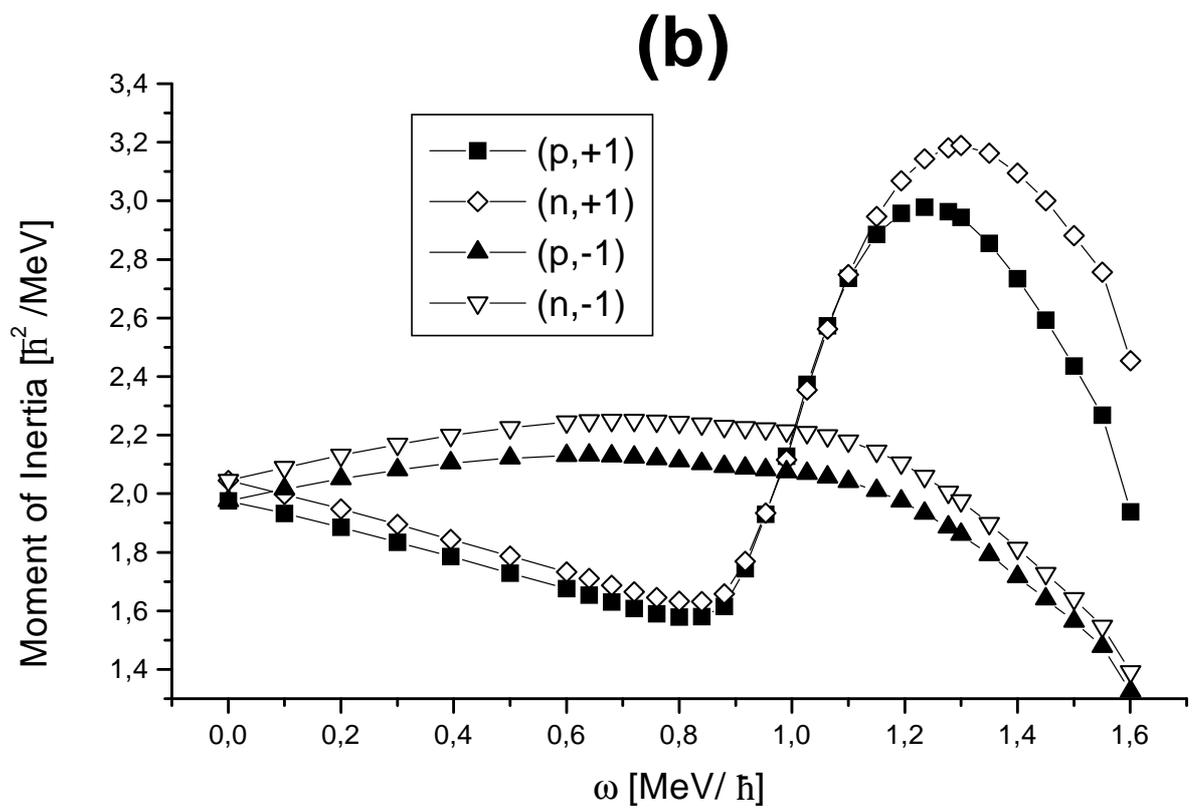
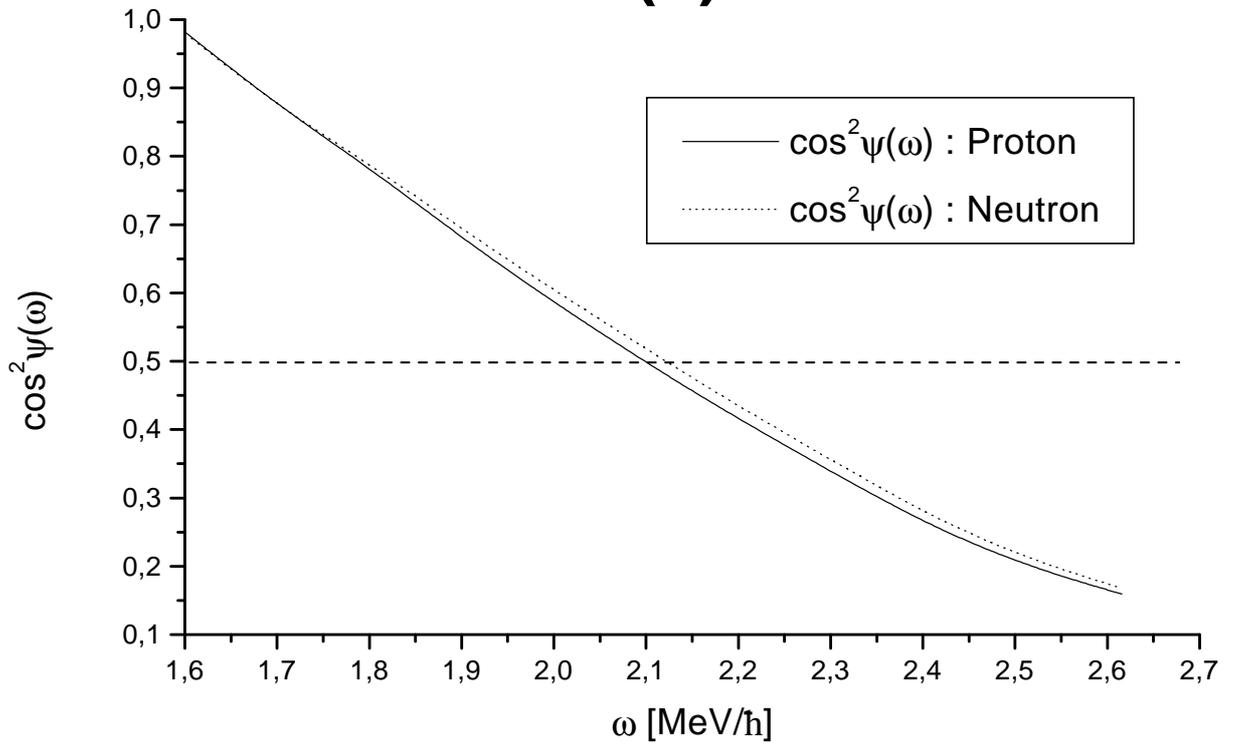
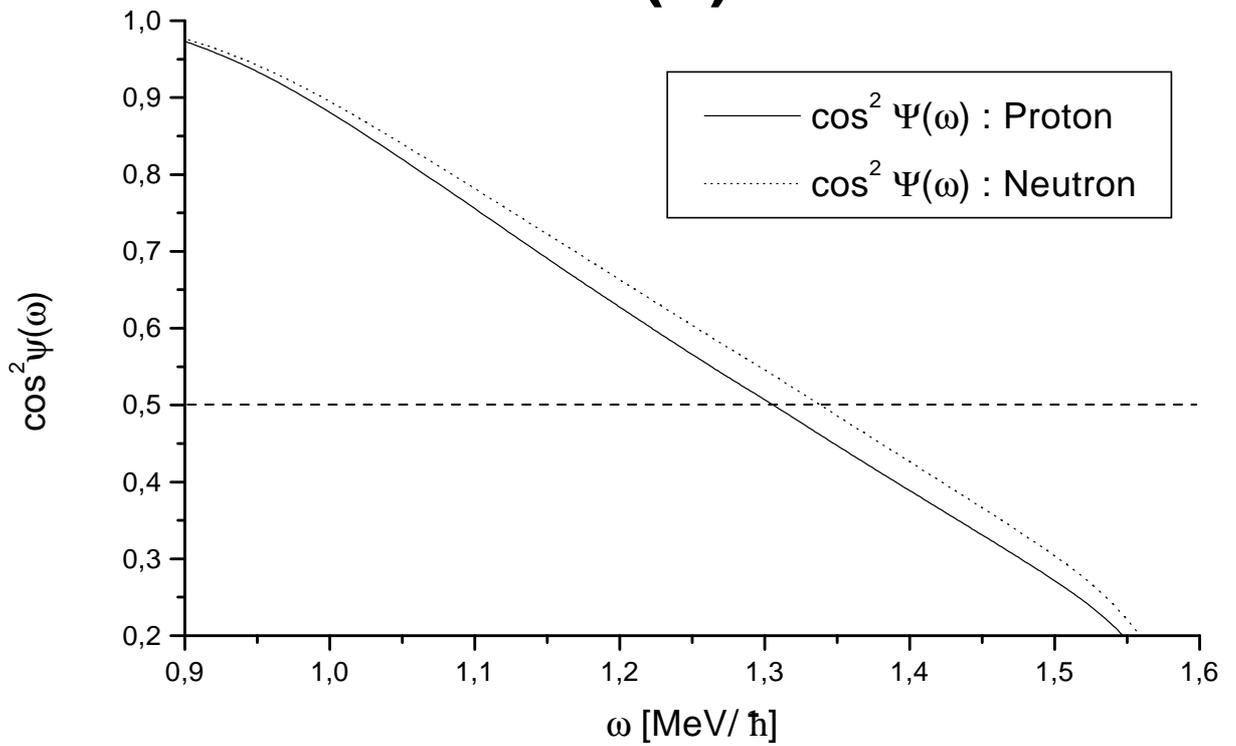
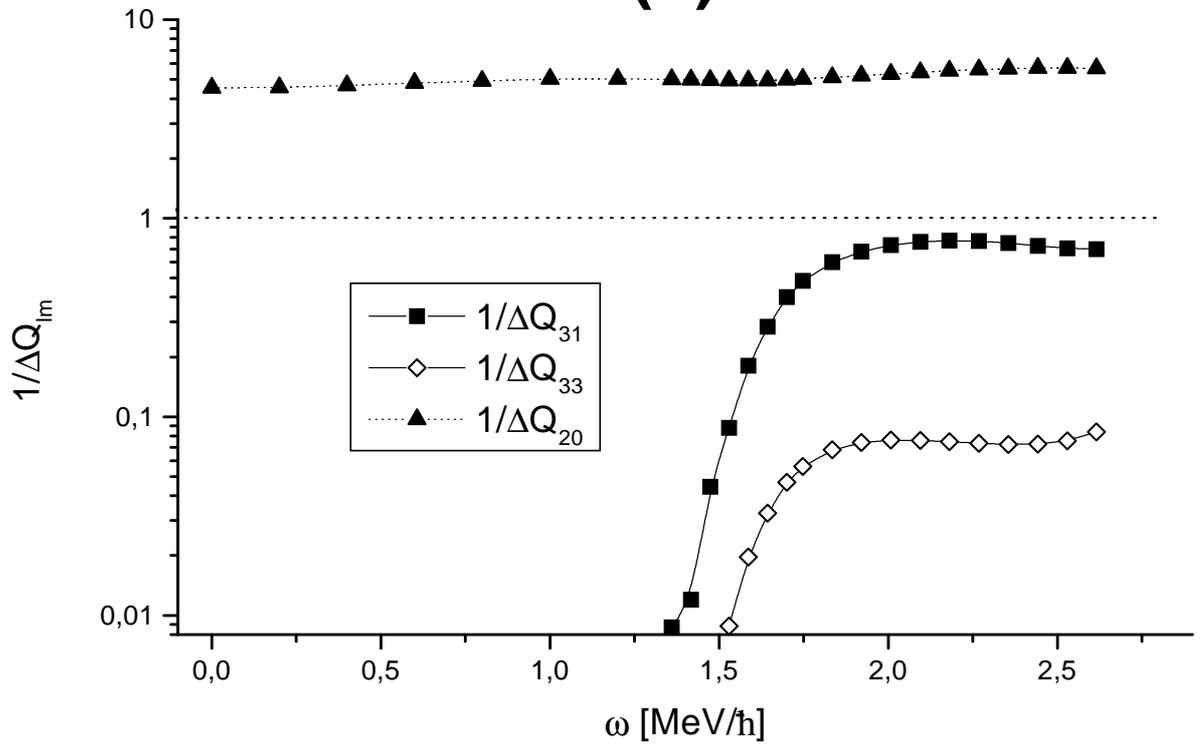
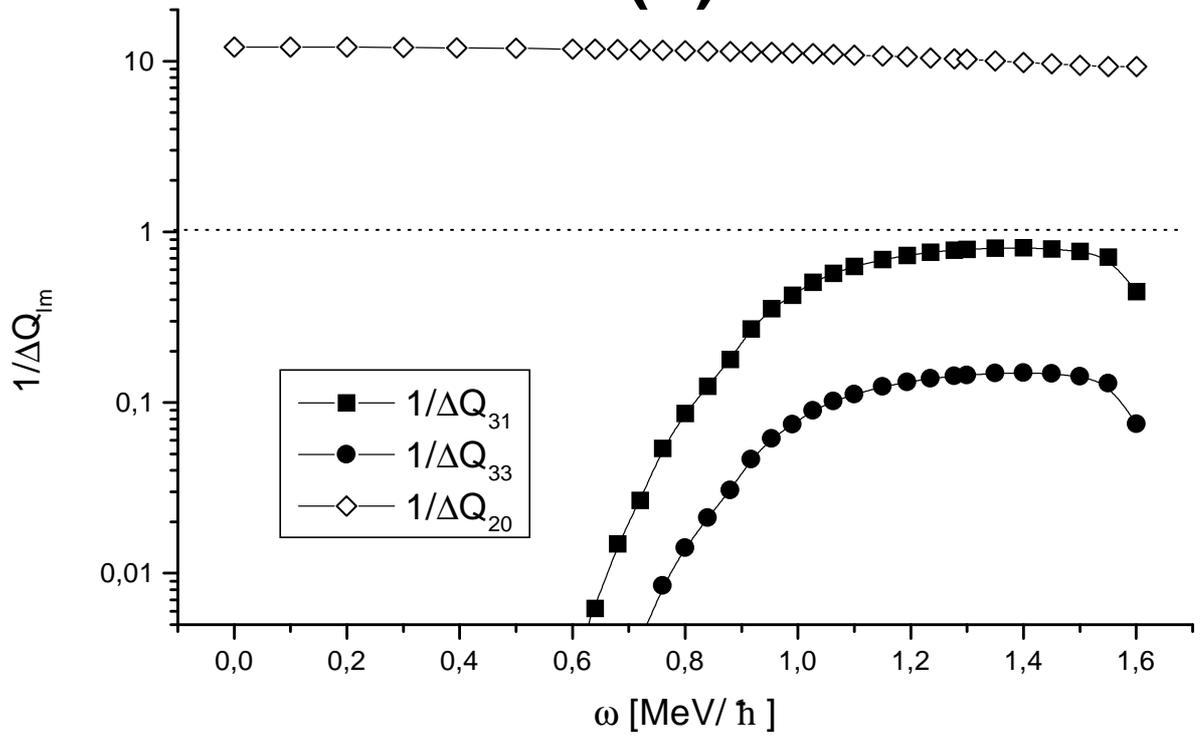
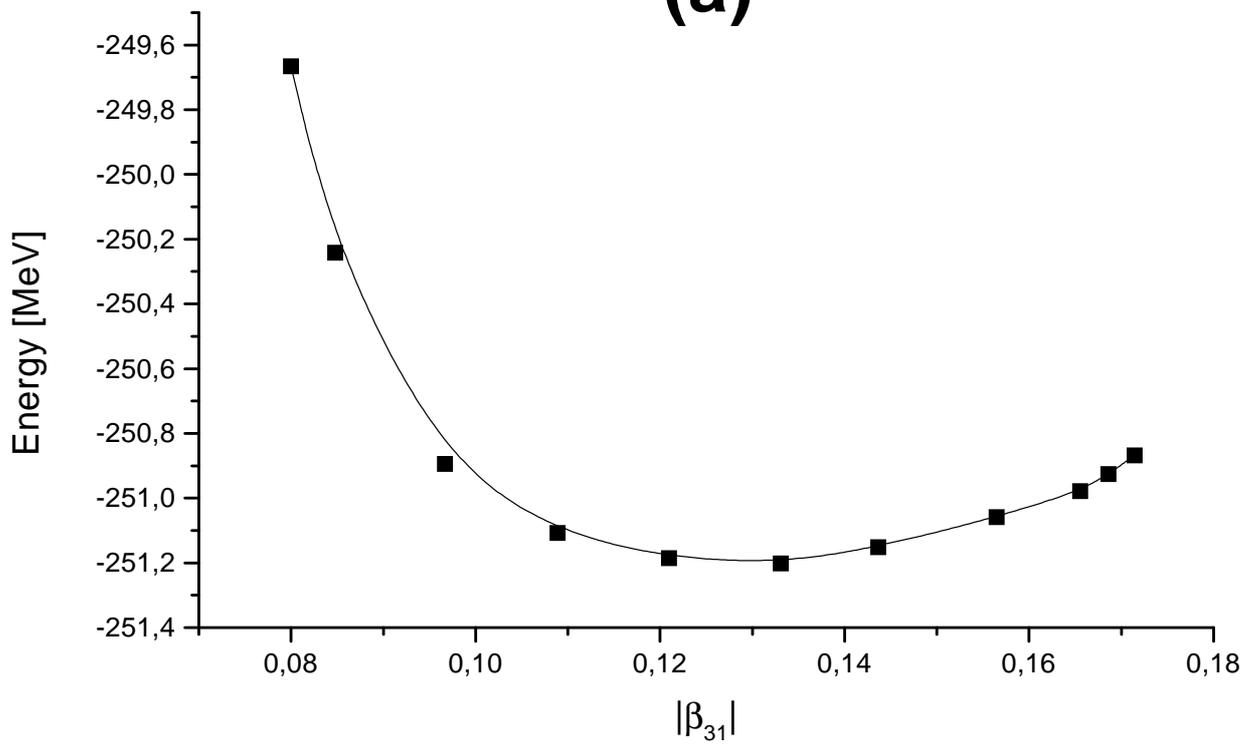
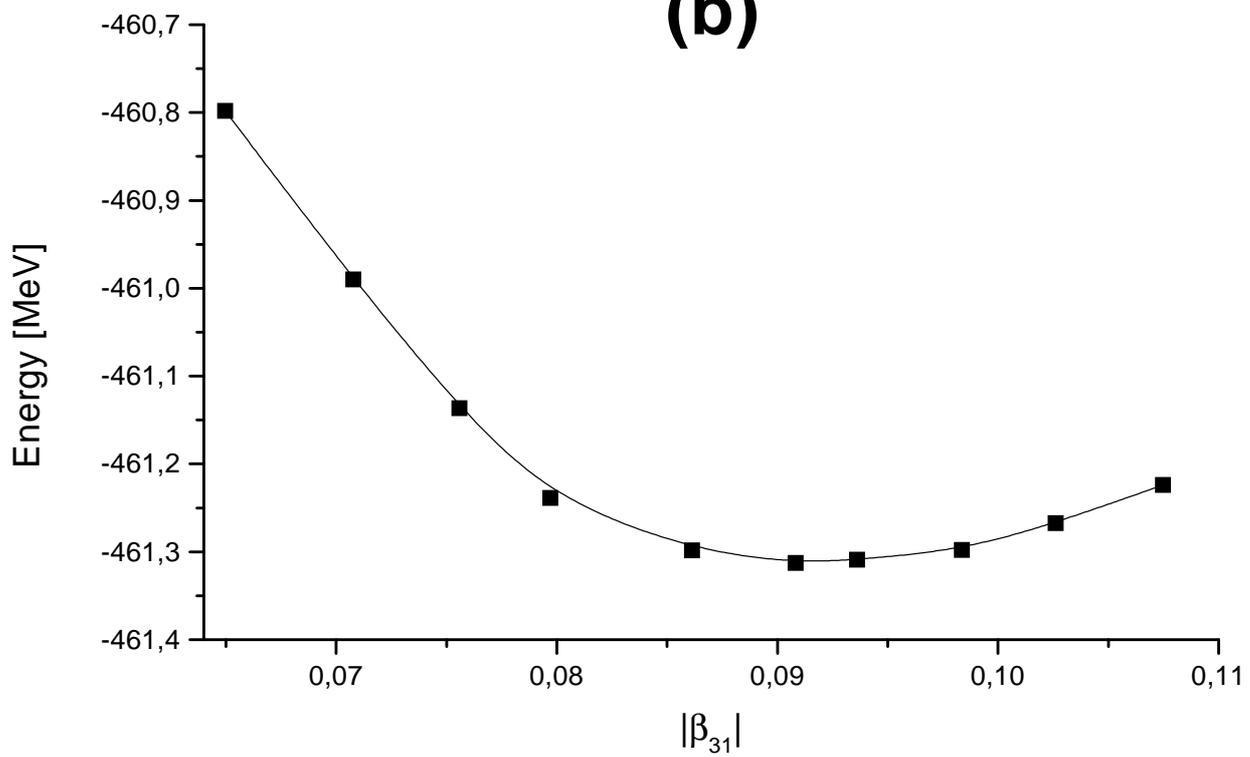
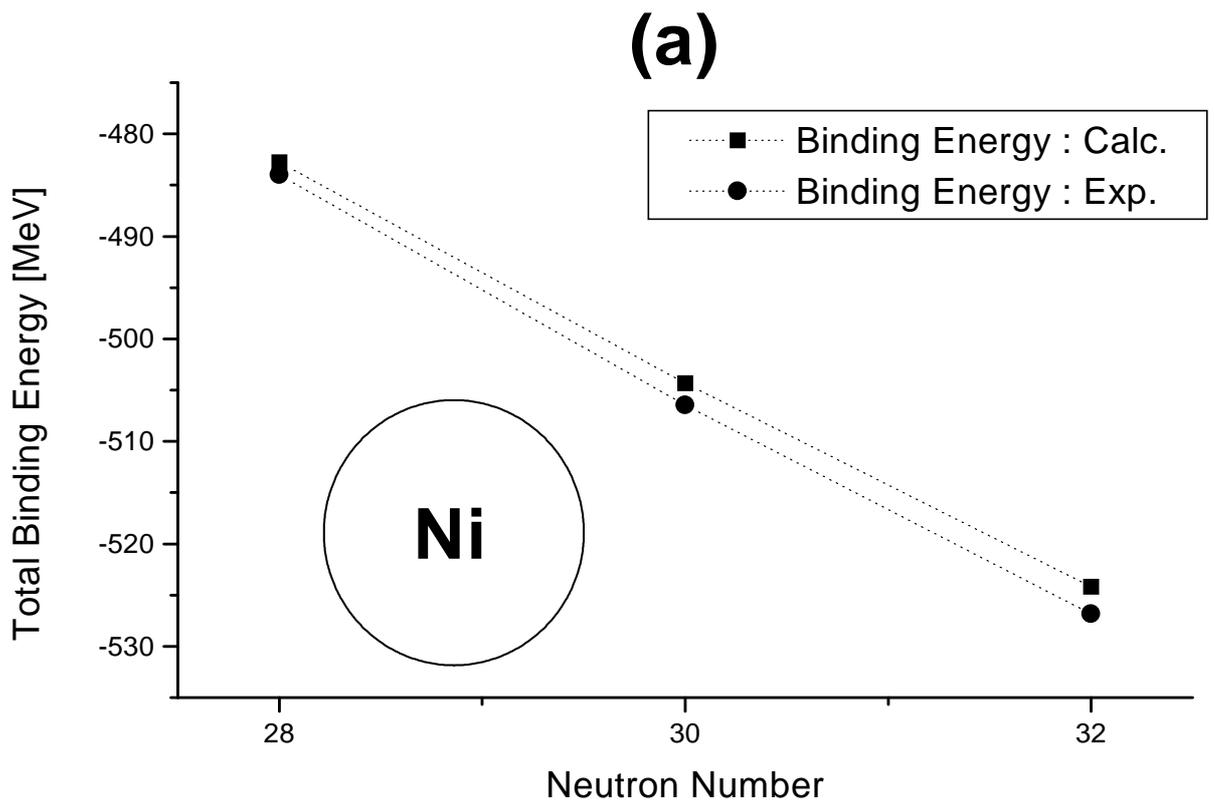
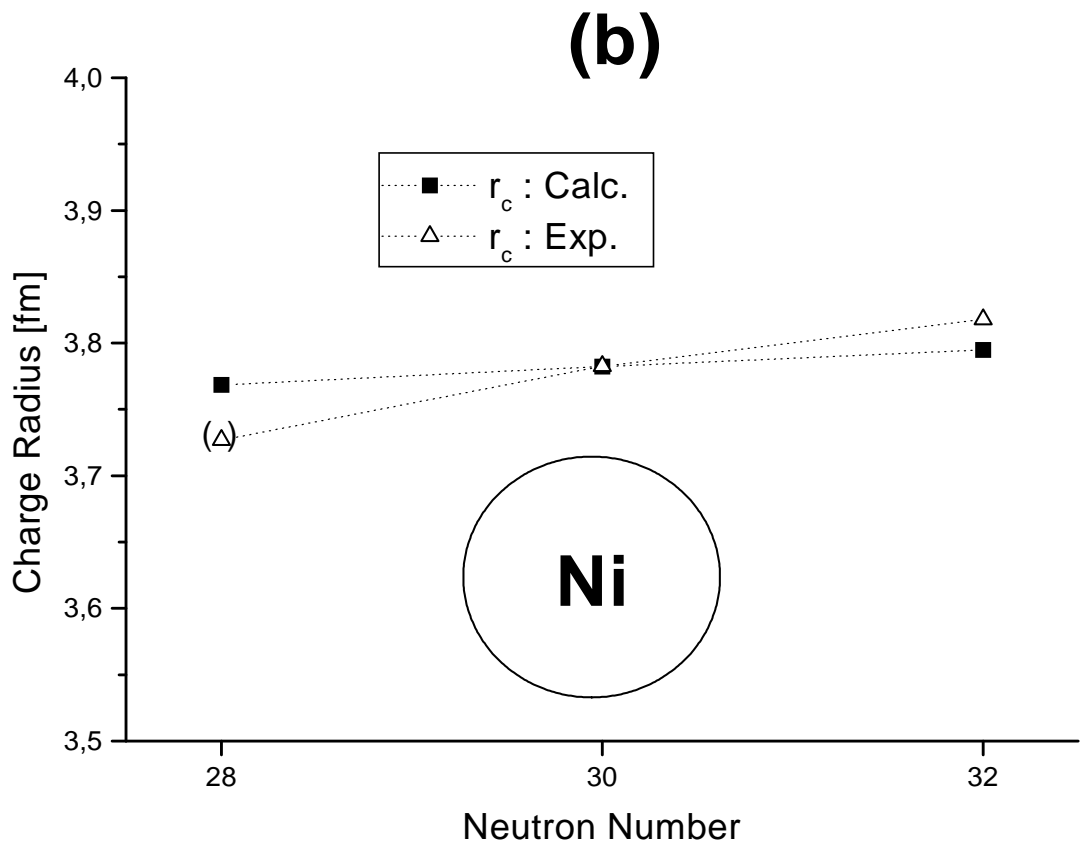
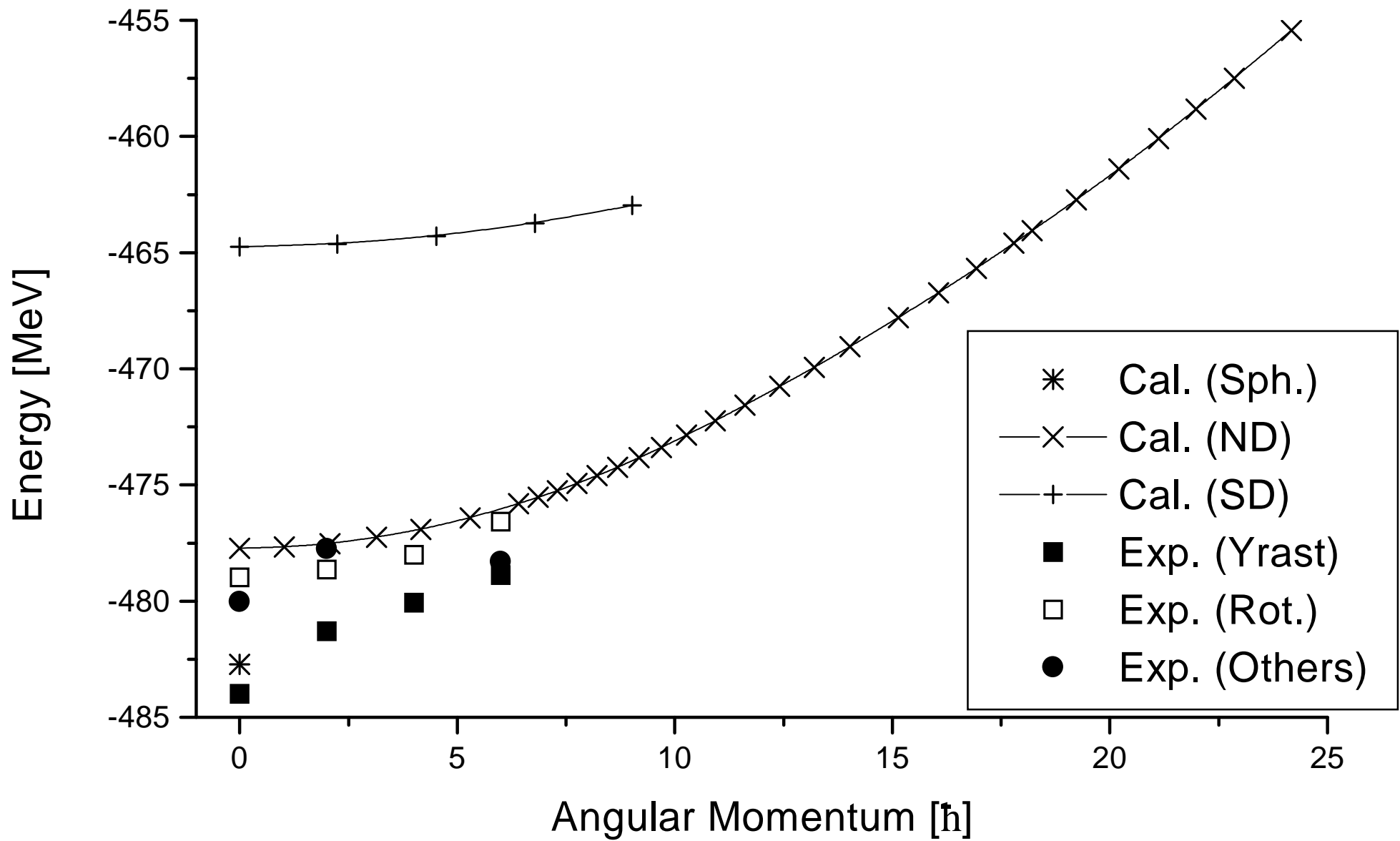
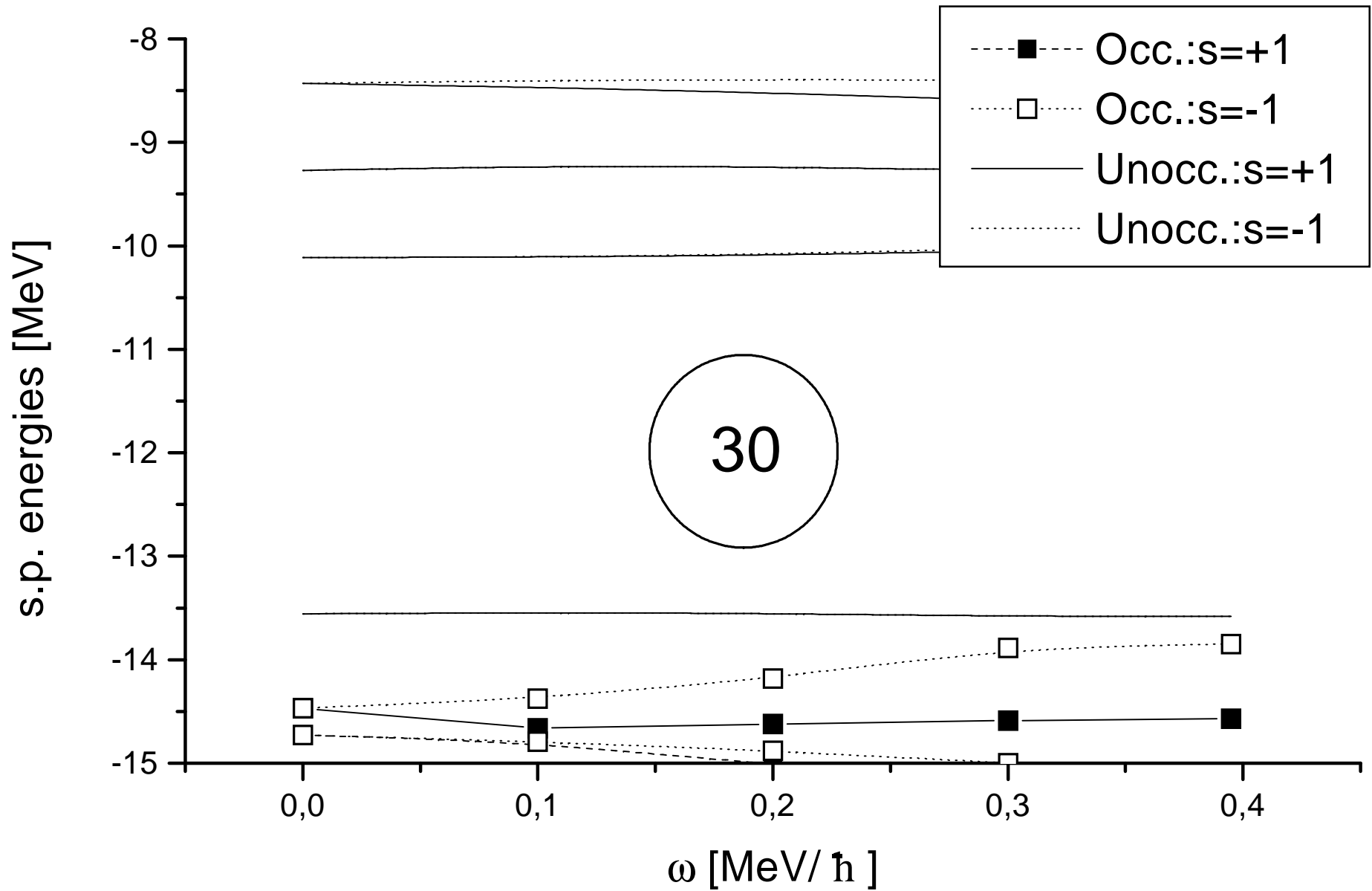
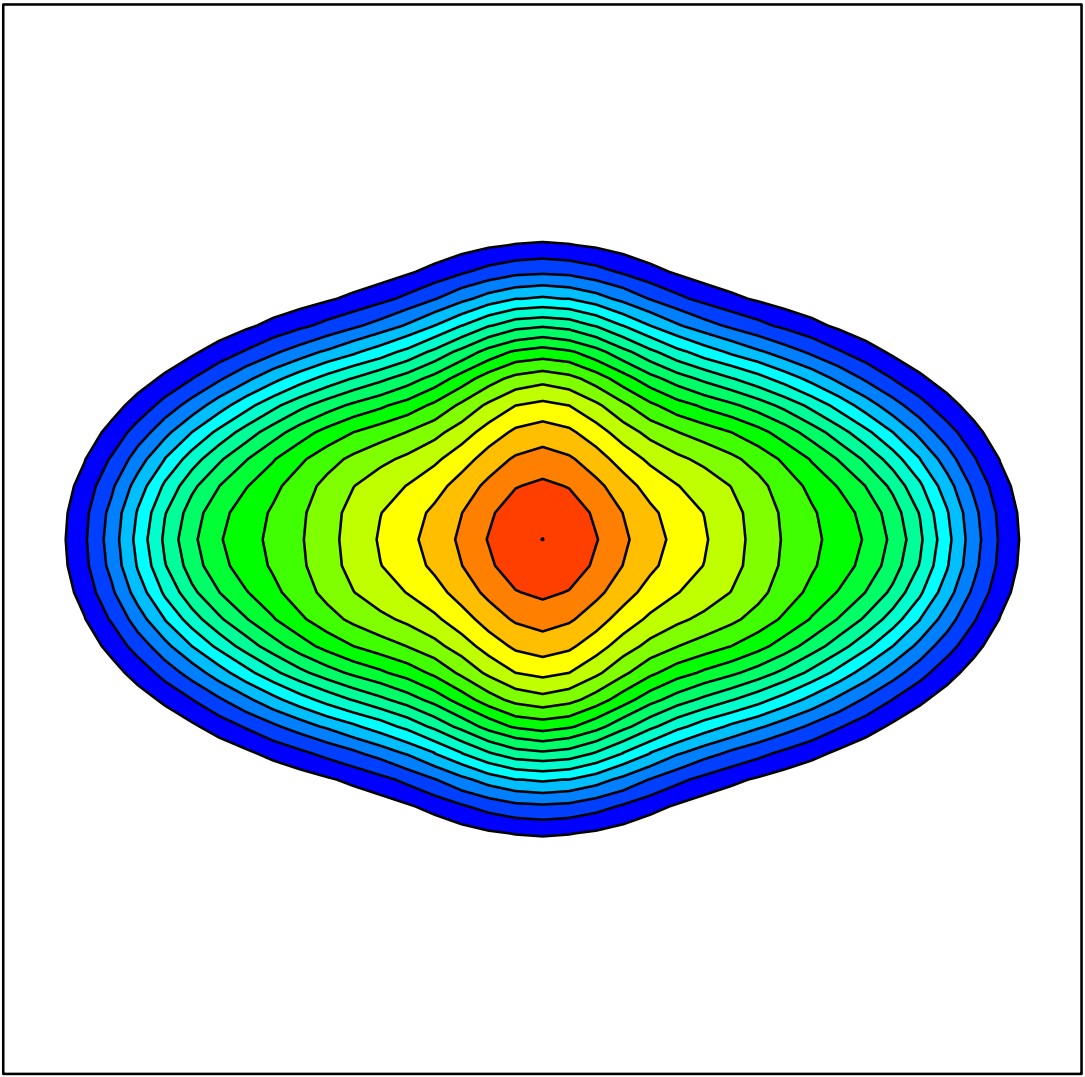
\end{document}